\documentclass[10pt]{article}
\usepackage{latexsym}
\usepackage{graphics,graphicx}
\usepackage{amsmath,amsfonts,amstext}
\usepackage{rotating,multirow}
\usepackage{hyperref}
\def\NO{\nonumber}

\newcommand{\be}{\begin{equation}}
\newcommand{\ee}{\end{equation}}

\def\bea{\begin{eqnarray}}
\def\eea{\end{eqnarray}}

\def\beqx{\begin{displaymath}}
\def\eeqx{\end{displaymath}}

\newcommand{\bmat}{\left(\begin{array}}
\newcommand{\emat}{\end{array}\right)}

\newtheorem{lemma}{Lemma}[section]

\def\a{\alpha}

\def\d{\delta}

\def\f{\phi}
\def\g{\gamma}

\def\k{\kappa}
\def\l{\lambda}
\def\m{\mu}
\def\n{\nu}

    \def\om{\omega}
\def\p{\pi}

    \def\th{\theta}
\def\r{\rho}
\def\s{\sigma}
\def\t{\tau}
\def\x{\xi}

\def\D{\Delta}
\def\F{\Phi}
\def\G{\Gamma}

    \def\Om{\Omega}

\def\S{\Sigma}

\def\vf{\varphi}

\def\cb{{\cal B}}
\def\cc{{\cal C}}

\def\ch{{\cal H}}

\def\cm{{\cal M}}
\def\cn{{\cal N}}
\def\co{{\cal O}}
\def\cp{{\cal P}}

\def\cs{{\cal S}}
\def\ct{{\cal T}}

\def\bo{{\raise-.3ex\hbox{\large$\Box$}}}               
\def\pa{\partial}                                       
\def\face{{\raise.2ex\hbox{$\displaystyle \bigodot$}\mskip-2.2mu \llap {$\ddot
        \smile$}}}                                   
\def\>{\rangle}                                      
\def\<{\langle}                                      
\def\hp{\widehat{\p}}                                

\newcommand{\sub}[1]{\phantom{}_{(#1)}\phantom{}}    
\newcommand{\super}[1]{\phantom{}^{(#1)}\phantom{}}  
\def\leftrightarrowfill{$\mathsurround=0pt \mathord\leftarrow \mkern-6mu
        \cleaders\hbox{$\mkern-2mu \mathord- \mkern-2mu$}\hfill
        \mkern-6mu \mathord\rightarrow$}        
\def\dvec#1{\vbox{\ialign{##\crcr
        \leftrightarrowfill\crcr\noalign{\kern-1pt\nointerlineskip}
        $\hfil\displaystyle{#1}\hfil$\crcr}}}           

\def\-{\hphantom{-}}

\title{\normalsize \mbox{ }\hspace{\fill}
\begin{minipage}{12 cm}{
\begin{flushright}
DESY-06-218\\
ZMP-HH/06-17\\
\tt hep-th/yymmnnn
\end{flushright}}{\hfill}
\end{minipage}
\\
{\Large\bf Multi-Trace Deformations in AdS/CFT: \\
Exploring the Vacuum Structure of the Deformed CFT}}

\author{\bf Ioannis Papadimitriou\footnote{ e-mail:
\tt ioannis.papadimitriou@desy.de } \\ \\
\em DESY Theory Group,\\ \em Notkestrasse 85, D-22603 Hamburg, Germany \\
\em and\\
\em Center for Mathematical Physics, \\
\em Bundesstrasse 55, D-20146 Hamburg, Germany.\\}

\date{}

\begin{document}
\maketitle

\renewcommand{\theequation}{\arabic{section}.\arabic{equation}}

\begin{abstract}

We present a general and systematic treatment of multi-trace deformations in the AdS/CFT
correspondence in the large N limit, pointing out and clarifying subtleties relating to the
formulation of the boundary value problem on a conformal boundary. We then apply this method
to study multi-trace deformations in the presence of a scalar VEV, which requires the coupling
to gravity to be taken into account. We show that supergravity solutions subject to `mixed'
boundary conditions are in one-to-one correspondence with critical points of the holographic
effective action of the dual theory in the presence of a multi-trace deformation, and
we find a number of new exact analytic solutions involving a minimally or conformally coupled
scalar field satisfying `mixed' boundary conditions. These include the generalization to any
dimension of the instanton solution recently found in hep-th/0611315. Finally, we provide a
systematic method for computing the holographic effective action in the presence of a multi-trace
deformation in a derivative expansion away from the conformal vacuum using Hamilton-Jacobi theory.
Requiring that this effective action exists and is bounded from below reproduces recent results on
the stability of the AdS vacuum in the presence of `mixed' boundary conditions.

\end{abstract}

\newpage

\tableofcontents
\addtocontents{toc}{\protect\setcounter{tocdepth}{2}}

\section{Introduction and summary of results}

Multi-trace deformations have been studied extensively in the context of the AdS/CFT correspondence
in the large $N$ limit, both classically \cite{Witten:2001ua,Berkooz:2002ug,Muck:2002gm,Minces:2002wp, Sever:2002fk, Aharony:2005sh, Elitzur:2005kz}
and at the one-loop level \cite{Gubser:2002vv,Hartman:2006dy,Diaz:2007an}. Most of this work, however, has focused on the
effect of multi-trace deformations on the conformal vacuum, in which case the back-reaction to the
geometry can be ignored. If the deforming operator though is allowed to acquire a non-zero VEV, then
the back-reaction can no longer be ignored and the coupling to gravity must be taken into account.
Only recently have multi-trace deformations in the presence of a scalar VEV been
considered, mainly in the context of Designer Gravity \cite{Henneaux:2002wm,Hertog:2004dr,Hertog:2004rz,
Hertog:2004jx,Hertog:2004bb,Hertog:2004ns,Hertog:2005hu,Hertog:2005hm,Amsel:2006uf,Hertog:2006wj,Amsel:2007im}.

In order for a CFT to admit multi-trace deformations it must contain operators with low enough
dimension. For double- or higher-trace deformations built out of a single operator, for example,
not to be irrelevant, the operator must have conformal dimension $\D\leq d/2$ in $d$ dimensions.
For scalar operators, for example, this means that the operator must have the `non-standard'
$\D_-$ dimension. This constraint, together with unitarity, which imposes a lower bound on the
dimension $\D$, severely restricts the CFTs admitting multi-trace deformations. The possibilities
are further narrowed if one insists that the undeformed CFT be supersymmetric. Since the AdS/CFT
dictionary relates multi-trace deformations in the large $N$ limit to a choice of boundary conditions
for the dual bulk supergravity fields \cite{Witten:2001ua}, these restrictions on the conformal
dimension of the operator translate into a condition on the mass of the dual supergravity fields
for them to admit the necessary generalized boundary conditions. We are then interested in gauged
supergravities that admit AdS vacua and have fields with mass close to the Breitenlohner-Freedman
bound \cite{Breitenlohner:1982bm}.

Both the maximal gauged supergravities in four and five dimensions contain scalars with the right mass, and
indeed black hole solutions with scalar hair that satisfy generalized boundary conditions were found
numerically in \cite{Hertog:2004dr}, following earlier work in three dimensions \cite{Henneaux:2002wm}.
Smooth instantons and gravitational soliton solutions of $\cn=8$ $D=4$ gauged supergravity with generalized
boundary conditions were also found numerically in \cite{Hertog:2004rz}, and shown to be related to
a Big Crunch geometry. More recently, exact solutions of $\cn=8$ $D=4$ gauged supergravity
obeying generalized boundary conditions were found analytically in \cite{Papadimitriou:2006dr} and
\cite{dHPP} and uplifted to eleven dimensions. The AdS/CFT identifies these solutions with
`vacua' or `states' in the dual {\em deformed} CFT. In particular, the extrema of the large $N$
quantum effective action for the VEV of the deforming operator are in one-to-one correspondence
with bulk solutions satisfying the relevant boundary conditions. These bulk solutions then
provide a window into the vacuum structure of the deformed theory.

A very interesting question, in particular, is whether the conformal vacuum - which generically
remains a vacuum of the deformed theory - is stable or not under certain boundary
conditions. The instantons found in \cite{Hertog:2004dr} and \cite{dHPP} show that it is not,
under the particular AdS-invariant boundary conditions that these instantons satisfy, since
these mediate the tunneling of the conformal vacuum to an instability region. This, of course,
does not contradict any of the well known stability theorems \cite{Gibbons:1983aq,Boucher:1984yx,
Townsend:1984iu}, because these apply only to certain special boundary conditions. The question
of stability with more general boundary conditions corresponding to multi-trace deformations
has been addressed recently in the context of Designer Gravity \cite{Hertog:2005hm,Amsel:2006uf,Hertog:2006wj,
Amsel:2007im}. The approach followed is a generalization of the spinorial argument of \cite{Witten:1981mf},
but as in the earlier work \cite{Boucher:1984yx,Townsend:1984iu} no supersymmetry is required.
The argument only relies on the existence `fake Killing spinors', which themselves can be constructed from a
`fake superpotential'. Non-perturbative stability then follows from the existence of a suitable
`fake superpotential'.

However, the AdS/CFT correspondence allows us to address the problem of non-perturbative stability
from a completely different point of view. Namely, if we knew the effective action of the dual
theory, then we would be able to address the question of stability/instability directly. We will show that the effective
action can be computed holographically in a derivative expansion using Hamilton-Jacobi theory
\cite{dBVV}. Requiring that this effective action exists and it is stable reproduces all known
stability results, including the recent results in the case of generalized boundary conditions.
This agreement can be traced to the fact that both arguments require global existence of a suitable
`fake superpotential'. In the latter case, however, this is interpreted as Hamilton's characteristic
function, which allows us to immediately generalize these results to other systems, such as
conformally coupled scalars.

The paper then is organized as follows. In Section \ref{mtds} we review a general
description of multi-trace deformations in the large $N$ limit, which relies on large $N$
factorization. This will make manifest the correspondence between multi-trace deformations
on the boundary and boundary conditions in the bulk in Section \ref{bvp}, where we
revisit the boundary value problem and the possible boundary conditions for the Klein-Gordon
operator in asymptotically locally AdS spaces. In particular, we present a general systematic
method to address multi-trace deformations and to properly account for the fact that the
boundary is a conformal boundary - as opposed to a hard boundary. As we show, this automatically
removes the divergences associated with the infinite volume of the space. Although we present
these results for scalar fields, they immediately generalize to any field admitting
boundary conditions corresponding to multi-trace deformations. In Section \ref{toy}
then we demonstrate the general method in the case of a free massive scalar field in a fixed AdS
background, reproducing in a concise way a number of known results. We then move on in
Section \ref{method} to include gravity and we describe in detail our method for computing the holographic
effective action of the dual theory in a systematic way based on Hamilton-Jacobi theory. This method
is then applied to the cases of scalars minimally and conformally coupled to gravity in Sections
\ref{minimal} and \ref{conformal} respectively, which contain our main results. In Section \ref{minimal}
we generalize the non-supersymmetric Poincar\'e domain wall solutions found in \cite{Papadimitriou:2006dr} to
arbitrary dimension, while the same is done for the instanton solution found in \cite{dHPP} in Section
\ref{conformal}. Moreover, we find {\em all} possible domain wall solutions - both flat and curved - for
the conformally coupled scalar in any dimension, and we show that this completely determines
the two-derivative effective action of the dual theory. Some technical results regarding the
variational problem for minimally and conformally coupled scalars, as well as the
Hamilton-Jacobi method for these systems, are collected in the Appendix.

\section{Multi-trace deformations in QFTs with a large N limit}
\label{mtds}
\setcounter{equation}{0}

In a quantum field theory with a standard large $N$ limit, large $N$ factorization allows for a
universal description of generic multi-trace deformations. As we now briefly review, the effect of
such a deformation can most naturally be described in terms of the generating functional
of the deforming operator and its Legendre transform \cite{Elitzur:2005kz}.

Let $\co(x)$ be a local, generically composite, gauge-invariant and single-trace operator transforming
in some representation of the relevant rank $N$ group. For concreteness we take this to be the adjoint
representation and we normalize the operator such that $\langle \co\rangle =O(N^0)$ as
$N\to\infty$. The dynamics of $\co(x)$ is encoded in the generating functional of connected correlators,
$W[J]$, which can be represented as a path integral over the fundamental degrees of freedom, $\{\f\}$, of
the theory, weighted by the action $S[\f]$, as
\be
e^{-W[J]}=\int [d\f]e^{-S[\f]- N^2\int d^dxJ(x)\co(x)}.
\ee
Since $W[J]$ is $O(N^2)$ as $N\to\infty$, it is convenient to work instead with  $w[J]\equiv N^{-2}W[J]$.
In particular, the one-point function of $\co(x)$ in the presence of a source is given by
\be
\s(x)\equiv \langle \co\rangle_J=\frac{\d w[J]}{\d J}.
\ee
Alternatively, the dynamics can be encoded in the Legendre transform of the generating functional,
$\G[\s]$, given by
\be
e^{-\G[\s]}=\int [dJ]e^{-N^2w[J]+N^2\int d^dx J(x)\s(x)}.
\ee
$\G[\s]$ is known as the effective action of the local operator $\co(x)$, or the
generating functional of 1PI diagrams. Again, it is useful to introduce the $O(N^0)$ quantity
$\bar{\G}[\s]=N^{-2}\G[\s]$, such that
\be\label{gap}
J(x)=-\frac{\d \bar{\G}[\s]}{\d \s}.
\ee

Suppose now that the action is deformed by a function, $f(\co)$, of the local operator
$\co(x)$ as $S_f[\f]=S[\f]+N^2\int d^dx f(\co)$. In the following we will only consider deformations for
which $f(0)=0$. The question we want to address now is how this deformation modifies the functionals $w[J]$ and
$\bar{\G}[\s]$. As we now show, large $N$ factorization allows for a very simple and universal answer
in the large-$N$ limit, which is summarized in Table \ref{multi-trace-defs}. Of course, beyond the
large $N$ approximation, the answer to this question is non-universal and much more involved, since the
operator $\co(x)$ will generically mix with other operators at $1/N$ order. We will only consider the leading
large $N$ behavior here.

Consider first the generating functional in the deformed theory, which is given by
\bea
e^{-N^2w_f[J_f]} & = & \int [d\f]e^{-S[\f]- N^2\int d^dx(J_f\co+f(\co))}\NO\\
& = & \int [d\f]e^{-S[\f]- N^2\int d^dx(J\co+f(\co)-f'(\s)\co)}\NO\\
&\stackrel{N\to\infty}{\approx}&e^{-N^2w[J]}e^{- N^2\int d^dx(f(\s)
-\s f'(\s))},
\eea
where we introduced $J\equiv J_f+f'(\s)$ in the second line in order to remove the linear
term from $f(\co)$ so that large $N$ factorization can be used in the last step.
This proves the result shown in the third row of Table \ref{multi-trace-defs}. Similarly,
the effective action in the deformed theory is given by
\bea
e^{-N^2\bar{\G}_f[\s]} & = & \int [dJ_f]e^{-N^2 w_f[J_f]+N^2\int d^dx J_f \s}
\NO\\
& \stackrel{N\to\infty}{\approx} & \int [dJ]e^{-N^2 w[J]}e^{-N^2\int d^dx (f(\s)-\s f'(\s))}
e^{N^2\int d^dx (J-f'(\s))\s}\NO\\
& = & e^{-N^2\bar{\G}[\s]-N^2\int d^dx f(\s)},
\eea
where we have used $[d J_f]=[dJ]$. This justifies the entry in the last row of Table \ref{multi-trace-defs}.
As we will review below, these universal results make manifest the fact that the AdS/CFT dictionary maps
multi-trace deformations of the boundary theory to a modification of the boundary conditions imposed
on the bulk fields. Before, however, we need to understand the boundary value problem for such bulk fields
in AdS.
\begin{table}
\begin{center}
\begin{tabular}{|c|c|c|}
\hline &&\\
  & Undeformed & Deformed \\ &&\\
\hline\hline &&\\
Source & $J$ & $J_f=J-f'(\s)$ \\  &&\\
VEV & $\s$ & $\s_f=\s$ \\  &&\\
Generating functional & $w[J]$ & $w_f[J_f]=w[J]+\int d^dx\left.\left(f(\s)-\s f'(\s)\right)
          \right|_{\s=\d w/\d J}$ \\ &&\\
Effective action & $\bar{\G}[\s]$ & $\bar{\G}_f[\s]=\bar{\G}[\s]+\int d^dxf(\s)$ \\ &&\\
\hline
\end{tabular}
\end{center}
\caption{The effect of a generic multi-trace deformation on the generating functional and on the effective
action in the large $N$ limit.}
\label{multi-trace-defs}
\end{table}

\section{The boundary value problem for the Klein-Gordon operator in AlAdS spaces}
\setcounter{equation}{0}
\label{bvp}

The gauge/gravity duality generically relates a theory of gravity in an asymptotically locally
anti de Sitter (AlAdS) space $\cm$ (see \cite{lectures} for a definition of an AlAdS space in the context
of the gauge/gravity duality) to a non-gravitational theory residing on the conformal
boundary $\pa\cm$ of $\cm$. Multi-trace deformations of the boundary theory are then related
to the choice of boundary conditions imposed on the bulk fields
\cite{Witten:2001ua,Berkooz:2002ug,Muck:2002gm,Minces:2002wp, Sever:2002fk,Gubser:2002vv, Aharony:2005sh, Elitzur:2005kz}.
The fact that the boundary of an AlAdS space is a {\em conformal} boundary, however,
demands some extra care when analyzing the boundary value problem. In particular, any rigorous
treatment should account for the following fact \cite{PS3}:
\begin{quote}
 {\em By the very definition of a `conformal boundary', any bulk field does not induce
a field on the boundary, but rather a field up to Weyl rescalings, i.e. a `conformal class'.
It follows that, in the absence of a conformal anomaly, any boundary condition must be imposed on
the conformal class and not on a conformal representative. In other words, any boundary
condition must be imposed on a `class function'. Although, this cannot be achieved if a conformal
anomaly is present, in that case one must ensure that the boundary condition is imposed on a quantity
that has a well defined transformation under Weyl rescalings.}
\end{quote}

This requirement, which we will make more precise and concrete below, has a number of important and
inevitable consequences that are often overlooked:
\begin{itemize}
 \item The well known boundary covariant counterterms {\em must} be added to the action
before one can study the variational problem and impose boundary conditions.

\item With the standard Dirichlet boundary conditions, the one-point function of an operator $\co(x)$
is in general {\em not} given by the normalizable mode of the corresponding bulk field. It is given
by the renormalized radial momentum \cite{PS1}. In general, the two differ by a local functional of
the non-normalizable mode, which is necessary to ensure that the Ward identities are fulfilled
\cite{dHSS,lectures}. In particular, it is the relation between the non-normalizable mode and the
renormalized radial momentum which is fundamentally related to the choice of boundary conditions
and {\em not} the relation between the non-normalizable and normalizable modes. Only when the two happen to
agree is one justified to use the normalizable mode instead of the renormalized momentum.

\end{itemize}
In view of these subtleties, and for the sake of completeness, we find it worthwhile to devote this
section to a careful and systematic analysis of the boundary value problem and to review how the AdS/CFT
dictionary relates the choice of boundary conditions for the bulk fields to multi-trace deformations of
the boundary theory. We also take this opportunity to spell out the formalism and notation which will be
used in the subsequent sections.

The metric on an AlAdS manifold, $\cm$ takes the form\footnote{We use Euclidean signature throughout this
paper.}
\be\label{gf-metric}
ds^2=dr^2+\g_{ij}(r,x)dx^idx^j,
\ee
where $\g_{ij}(r,x)\sim e^{2r/l}g\sub{0}_{ij}(x)$ as $r\to\infty$ and hence,
the conformal boundary, $\pa\cm$, is located at $r=\infty$. The metric
$g\sub{0}_{ij}(x)$ is a metric on the conformal boundary, or more precisely, a {\em representative}
of the conformal class of boundary metrics. AlAdS metrics arise naturally
as solutions of Einstein's equations with a negative cosmological constant,
possibly including matter whose stress tensor falls fast enough asymptotically \cite{lectures}.

To set up the formalism, we will study the simplest possible boundary value problem
on the background of such a manifold, namely that of the Klein-Gordon equation for a
scalar field,
\be\label{K-G}
\left(-\square_g+m^2\right)\f=0.
\ee
One can include interactions in this equation, and we will do so later on, but these
are irrelevant for the boundary value problem as long as they do not modify the asymptotic
form of the metric. Any solution of (\ref{K-G}) in the background (\ref{gf-metric})
takes the asymptotic form
\be\label{asymp_exp}
\f\sim \left\{\begin{matrix}
e^{-\D_-r/l}(\f_-(x)+\cdots)+ e^{-\D_+r/l}(\f_+(x)+\cdots),& m^2l^2>-(d/2)^2,\\
\\
e^{-dr/2l}\frac{r}{l}(\f_-(x)+\cdots)+e^{-dr/2l}(\f_+(x)+\cdots), & m^2l^2=-(d/2)^2,\\
\end{matrix}\right.
\ee
where $\D_\pm$, $\D_+\geq \D_-$, are the roots of the equation
$m^2l^2=\D(\D-d)$, and the functions $\f_-(x)$ and $\f_+(x)$, known respectively
as the non-normalizable and normalizable modes, are totally
arbitrary functions of the transverse coordinates, $x^i$. A boundary
condition amounts to a choice of a function $J(\f_-,\f_+)$ of the two modes
that is kept constant on the boundary, thus reducing by half the degrees of freedom.
The best known example is that of Dirichlet boundary conditions, where one
chooses $J(\f_-,\f_+)=\f_-$, and so the only degree of freedom remaining is the
normalizable mode $\f_+$.

We could now try to study boundary conditions by considering different choices of the function
$J(\f_-,\f_+)$, which is in fact what has been done in the vast majority of the literature on
the subject. Although, this approach happens to work in certain cases, generically it is fundamentally
problematic for two closely related reasons. Firstly, the quantity $\f_+$, contrary to
the non-normalizable mode $\f_-$, has no well defined transformation under Weyl rescalings and
hence it is an ill defined quantity from the boundary point of view. Secondly, precisely
because $\f_+$ is not well defined on the boundary, the function $J(\f_-,\f_+)$ one
would use to define the boundary condition does {\em not} have a direct meaning (as a source)
on the boundary.  As we mentioned above and we will now explain in detail,
both problems are resolved if one replaces $\f_+$ in this analysis with the renormalized
radial momentum \cite{PS1}, which does have a definite transformation under
Weyl rescalings and hence it is a well defined boundary quantity, like $\f_-$. The
renormalized radial momentum in general differs from $\f_+$ by a local functional
of $\f_-$, which is essential to ensure that the Ward identities are satisfied \cite{lectures}.
It is precisely these local terms that make the renormalized momentum have a well defined
transformation rule under Weyl rescalings. Unless these local terms {\em happen}
to vanish, and one needs to demonstrate that they do, we are forced to use
the radial Hamiltonian formulation in order to discuss generalized boundary
conditions consistently.

\subsection{The variational problem in the presence of a conformal boundary}

Since the conformal boundary, $\pa\cm$, is located at infinity, we need to introduce a regulating
surface, $\S_r$, diffeomorphic to the boundary, but at a finite value of the radial coordinate
$r$. One then formulates the variational problem on $\S_r$ and in the end the regulator is removed
by sending $r\to \infty$. It is crucial, however, to keep in mind that the conformal boundary
$\pa\cm$ and the hard boundary introduced by the regulating surface $\S_r$ are very different in nature.
In particular, the regulating surface breaks explicitly the invariance under Weyl rescalings that
the conformal boundary possesses. It follows that not any variational problem that makes
sense on $\S_r$ will make sense as the regulator is removed. It will only make sense
provided the variational problem on $\S_r$ is formulated in terms of conformal class functions.
Before we discuss how this can be achieved, though, let us consider the general variational
problem on the regulating surface $\S_r$.

Given an action $S[\f]$ on a space $\cm_r$ with a boundary $\S_r=\pa\cm_r$, one is naturally led
to the radial Hamiltonian formulation of the bulk dynamics by considering the variational problem
for the action $S[\f]$. Indeed, a generic variation of the bulk action with respect
to the scalar field generates a boundary term of the form\footnote{The complete expressions for
the variation of the action when the scalar is minimally or conformally coupled to gravity are presented in
Appendix \ref{appendix}.}
\be\label{var}
\d_\f S = \int_{\S_r}d^d x\p_\f\d\f,
\ee
where $\p_\f$ is the canonical momentum conjugate to $\f$ and the Hamiltonian `time' is taken to
be the radial coordinate $r$ orthogonal to the boundary $\S_r$. For the bulk action giving
the Klein-Gordon equation as the equation of motion, the canonical momentum is simply
$\p_\f=\sqrt{\g}\dot{\f}$, where the dot denotes a derivative with respect to the radial coordinate, $r$.
Were $\S_r$ the true boundary, we could impose any boundary condition
compatible with the variation (\ref{var}). But we actually have to send $r\to\infty$ in the end,
and the integrand in (\ref{var}) does not have a well defined transformation
under shifts in $r$. Hence, if we impose a boundary condition on (\ref{var}), we will not be able to
`push' this boundary condition to the true boundary at $r\to \infty$. What we need to do first, is to
find a covariant way of modifying (\ref{var}), without changing the bulk dynamics of course, such
that the result has a well defined transformation - in fact, remains {\em invariant} - under radial
shifts. A systematic way of constructing quantities which are both covariant with respect to $\S_r$
diffeomorphisms and have a well defined transformation under radial shifts in the vicinity of the
conformal boundary $\pa\cm$, is based on the following observation \cite{PS1}: The asymptotic form
$\f\sim e^{-\D_-r/l}\f_-(x)$ of the scalar field allows us to write the radial derivative in the
from\footnote{\label{comment}In general, one must sum over all induced fields on $\S_r$,
including the induced metric, $\g_{ij}$. In particular, the dilatation operator contains the term
$\int_{\S_r} d^dx2\g_{ij}\frac{\d}{\d\g_{ij}}$ and so it acts on the volume element on $\S_r$ as
$\d_D\sqrt{\g}=d\sqrt{\g}$.}
\be\label{rad-der-exp}
\pa_r=\int_{\S_r} d^dx\dot{\f}\frac{\d}{\d\f}\sim \frac1l\d_D,
\ee
where
\be\label{dilatation}
\d_D=-\D_-\int_{\S_r} d^dx\f\frac{\d}{\d\f},
\ee
is the dilatation operator, and $\sim$ means that only the leading asymptotic behavior as $r\to\infty$ is shown.
It follows that quantities that have a well defined transformation under radial shifts correspond to
eigenfunctions of the dilatation operator. However, by trading the radial derivative for the dilatation operator
we also automatically achieve covariance with respect to $\S_r$ diffeomorphisms. The dilatation operator
(\ref{dilatation}), therefore, provides us with a way to decompose the integrand in (\ref{var}),
which does not transform in a controlled way under radial translations, into pieces with a well defined
transformation.

This is achieved by expanding the canonical momentum $\p_\f$ in eigenfunctions of the dilatation
operator as
\be\label{mom-exp}
\p_\f=\sqrt{\g}\left(\p\sub{\D_-}+\cdots +\p\sub{\D_+}+\cdots\right),
\ee
where $\d_D\p\sub{n}=-n\p\sub{n}$ for all $n$.\footnote{A logarithmic term should be included in general to
account for a possible conformal anomaly. In the presence of such an anomaly, $\p\sub{\D_+}$ transforms
inhomogeneously under the dilatation operator \cite{PS1}.} This is simply a formal expansion at this point, as
is (\ref{asymp_exp}), but the fact that $\f$ and $\p_\f$ do admit the expansions (\ref{asymp_exp}) and
(\ref{mom-exp}) respectively, is a consequence of the equation of motion. Since $\p_\f=\sqrt{\g}\dot{\f}$,
one can insert the expansion (\ref{mom-exp}) in (\ref{rad-der-exp}) to obtain a formal expansion of the
radial derivative in covariant functional operators of definite dilatation weight. Substituting this expansion
for the radial derivative, together with the momentum expansion (\ref{mom-exp}), into the equation of motion (\ref{K-G})
and matching terms of equal dilatation weight one can determine iteratively all terms $\p\sub{n}$ for $n<\D_+$
as {\em local} functionals of $\f$. The fact that these terms turn out to be local functionals of the induced
field $\f$ is the crucial ingredient which allows us to formulate the boundary value problem on the conformal
boundary. In particular, we can write
\be\label{def-counterterms}
\int_{\S_r}d^d x\sqrt{\g}\sum_{n<\D_+}\p\sub{n}\d\f=-\d S_{\rm ct}[\f],
\ee
where $S_{\rm ct}[\f]$ is a {\em local} functional of $\f$.
From (\ref{var}) then follows that if the local functional $S_{\rm ct}$ on $\S_r$
is added to the bulk action on $\cm_r$, then a generic variation of the total action
produces the boundary term\footnote{Terms of higher dilatation weight drop out in the limit $r\to\infty$.}
\be\label{bt}
\d(S+S_{\rm ct})=\int_{\S_r}d^d x\sqrt{\g}\p\sub{\D_+}\d\f.
\ee
Even though this might seem little different from the original expression (\ref{var}),
the difference is in fact fundamental: contrary to (\ref{var}) the integrand in
(\ref{bt}) is invariant under radial translations since
$\d_D(\sqrt{\g}\p\sub{\D_+}\d\f)=(d-\D_+-\D_-)\sqrt{\g}\p\sub{\D_+}\d\f=0$,
where we have used $\d_D\sqrt{\g}=d\sqrt{\g}$ (see footnote \ref{comment}).
If follows that we can now send $r\to\infty$ and any boundary condition
formulated in terms of $\f$ and the renormalized momentum $\p\sub{\D_+}$
will remain unchanged and meaningful in this limit.

Two comments are in order here. First, note that the counterterms
we have defined via (\ref{def-counterterms}), and which were
introduced only on the basis that they are required to make the
variational problem on the conformal boundary well posed, are
{\em identical} with the standard boundary counterterms
that are traditionally added to make the on-shell action finite.
Indeed, the fact that (\ref{bt}) has a finite limit as
$r\to \infty$ implies that the renormalized on-shell action
$S_{\rm ren}\equiv(S+S_{\rm ct})$, remains finite as the regulator
is removed. The same local counterterms are therefore required to make
the variational problem well posed and to remove the infra red divergences
of the on-shell action. We would like to view the former, however, as the more
fundamental property. Indeed, the divergences of the on-shell action are
merely a manifestation of the fact that the variational problem is
not formulated properly \cite{PS3}. Of course, there is as usual a freedom of adding extra finite local terms
to the counterterms $S_{\rm ct}$. In the case of Dirichlet boundary conditions this is the well known
renormalization scheme dependence. As we will see below, however, the interpretation of this
freedom in the dual field theory crucially depends on the boundary conditions. In particular, for boundary conditions
other than Dirichlet, it {\em does not} correspond to a renormalization scheme dependence.  

The second comment concerns some notation. It is very useful to introduce
\be
\hp\sub{\D_+}\equiv\left\{\begin{matrix} \lim_{r\to\infty}e^{\D_+r/l}\p\sub{\D_+}, & \D_+>d/2, \\ &\\
                           \lim_{r\to\infty}r^2e^{dr/2l}\p\sub{\D_+}, &  \D_+=d/2,
                          \end{matrix}\right.
\ee
which allows us to explicitly evaluate the limit $r\to\infty$ in (\ref{bt}) as
\be\label{bt-lim}
\framebox[2.5in]{
\begin{minipage}{3.0in}
\hspace{.2in}
\begin{equation*}
\d(S+S_{\rm ct})=\int_{\pa\cm}d^d x\sqrt{g\sub{0}}\widehat{\p}\sub{\D_+}\d\f_-.
\end{equation*}
\end{minipage}
}
\ee
The boundary value problem on the conformal boundary is then naturally formulated
in terms of the two modes $\f_-(x)$ and $\widehat{\p}\sub{\D_+}$. Comparing
the expansions (\ref{asymp_exp}) and (\ref{mom-exp}), e.g. for $\D_-\neq\D_+$, one finds that
$\hp\sub{\D_+}=-(\D_+-\D_-)\f_{+}(x)/l+C[\f_-(x)]$, where $C[\f_-(x)]$ is a local functional of
$\f_-(x)$ depending on the space dimension as well as on the bulk dynamics.\footnote{We will see below that
in the case of Dirichlet boundary conditions $\hp\sub{\D_+}$ is identified via the AdS/CFT dictionary with the
one-point function of the dual operator in the presence of an arbitrary source $\f_-(x)$. The fact that
the one-point function generically contains a non-linear but local functional, $C[\f_-(x)]$, of the source was
shown originally in \cite{dHSS}.} Interestingly, as we will later show, for the boundary conditions relevant to
multi-trace deformations it turns out that $C[\f_-(x)]$ vanishes identically - thus a posteriori
justifying the use of $\f_+(x)$ instead of the renormalized momentum in the literature. However, in general, it is
$\hp\sub{\D_+}$ and not the normalizable mode which has a well defined transformation under boundary Weyl transformations.

\subsection{Boundary conditions}

The expression (\ref{bt-lim}) is our starting point for studying the
possible boundary conditions on the conformal boundary.
A boundary condition is in general a choice of a function, $J(\f_-,\hp\sub{\D_+})$,
of the two independent modes, $\f_-$ and $\hp\sub{\D_+}$, that is kept fixed on the boundary.
Note that we have now replaced $\f_+$ with $\hp\sub{\D_+}$, which as we discussed, is
necessary in order for the boundary condition to be well defined on the
conformal boundary. In order to impose the boundary condition $\d J(\f_-,\hp\sub{\D_+})=0$,
we need to add a suitable (finite) boundary term, $S_J[\f_-,\hp\sub{\D_+}]$, to the action such
that\footnote{The apparently alternative boundary condition $B_J(\f_-,\hp\sub{\D_+})=0$,
is not acceptable in the context of the AdS/CFT correspondence. The reason is that such a
boundary condition really reduces by half the degrees of freedom. In AdS/CFT, however, the
boundary condition does halve the bulk degrees of freedom, but the lost half reappears as a
source on the boundary.}
\be
\d(S+S_{\rm ct}+S_J)=\int_{\pa\cm}d^dx\sqrt{g\sub{0}}B_J(\f_-,\hp\sub{\D_+})\d J(\f_-,\hp\sub{\D_+}),
\ee
where $B_J(\f_-,\hp\sub{\D_+})$ is some function that depends on the choice of $J(\f_-,\hp\sub{\D_+})$.
\begin{table}
\begin{center}
\begin{tabular}{|c|c|c|}
\hline &&\\
  & $J(\f_-,\hp\sub{\D_+})$ & $S_J[\f_-,\hp\sub{\D_+}]$ \\ &&\\
\hline\hline &&\\
Dirichlet & $J_+=\f_-$ & $S_+=0$ \\  &&\\
Neumann & $J_-= -\hp\sub{\D_+}$ & $S_-=-\int_{\pa\cm} d^dx\sqrt{g\sub{0}}\f_-\hp\sub{\D_+}$ \\  &&\\
Mixed   & $J_{f_-}=-\hp\sub{\D_+}-f'(\f_-)$ & $S_{f_-}=S_-+\int_{\pa\cm}
d^dx\sqrt{g\sub{0}}(f(\f_-)-\f_- f'(\f_-))$ \\ &&\\
\hline
\end{tabular}
\end{center}
\caption{The three inequivalent boundary conditions for a scalar field in an AlAdS background, along with
the corresponding boundary terms required to impose them. Notice that the Neumann boundary condition is a
special case of the Mixed boundary condition, obtained by choosing the function $f(\f_-)$ to be identically zero.}
\label{bcs}
\end{table}

A physical solution of the equations of motion, subject to the boundary condition defined by $J(\f_-,\hp\sub{\D_+})$,
satisfies $J(\f_-,\hp\sub{\D_+})=0$. Note that this definition of `physical
solutions' excludes solutions that describe {\em single}-trace deformations, which require
a non-zero source. It follows that there are two qualitatively different universality classes of possible
boundary conditions, depending on whether the mode $\f_-(x)$ in the corresponding physical solutions is zero
or not, which lead to different leading asymptotics for the physical solutions. $\f_-(x)$ is zero in the physical
solutions provided the source $J(\f_-,\hp\sub{\D_+})$ is a function of $\f_-$ only. When $\f_-(x)$ is non-zero in
the physical solutions, then the relation $J(\f_-,\hp\sub{\D_+})=0$ determines $\hp\sub{\D_+}$ as a function of
$\f_-(x)$. The three inequivalent choices of boundary conditions, along with the corresponding boundary term,
$S_J[\f_-,\hp\sub{\D_+}]$, that should be added to the action are listed in Table \ref{bcs}.\footnote{These
boundary conditions exhaust all possible relations between $\hp\sub{\D_+}$ and $\f_-(x)$ in the physical
solutions, and so all possible Hilbert spaces obtained by quantizing the scalar field. Nevertheless, there
is an apparent redundancy in the choice of the source $J(\f_-,\hp\sub{\D_+})$. For example, $J=\f_-$ leads
to the same Hilbert space as $J=\f_-^2$. In the context of the AdS/CFT correspondence, this redundancy is
mapped to an analogous redundancy in defining the generating functional of a given operator, and in particular
in the choice of its source. Table \ref{bcs} shows the standard minimal choices.}

However, requiring that the (static) solutions subject to the boundary conditions in Table \ref{bcs}
are perturbatively stable imposes restrictions on the allowed values of the scalar mass $m^2$. In particular,
for Dirichlet boundary conditions, stability requires that the scalar mass satisfies the Breitenlohner-Freedman
(BF) bound \cite{Breitenlohner:1982bm}, $m^2l^2\geq -(d/2)^2$, while if $\f_-(x)\neq 0$ in the solution, i.e.
for Neumann and Mixed boundary conditions, stability requires that the mass squared is in the range
\cite{Breitenlohner:1982bm,Balasubramanian:1998sn,Klebanov:1999tb}
\be\label{mass_range}
-\left(\frac{d}{2}\right)^2\leq m^2l^2\leq -\left(\frac{d}{2}\right)^2+1.
\ee
We will later show that these stability conditions follow immediately from the requirement
that there exists a stable holographic effective potential for the dual operator.
Moreover, generically only one boundary condition will be consistent with supersymmetry once the scalar
is embedded in some gauged supergravity \cite{Breitenlohner:1982bm,Breitenlohner:1982jf}.

The mass constraint (\ref{mass_range}) for the Neumann or Mixed boundary conditions to be admissible has
a remarkable and somewhat surprising consequence. Namely, it ensures that the local functional, $C[\f_-(x)]$,
which, as we pointed out above, distinguishes in general the renormalized momentum $\hp\sub{\D_+}$ from the
normalizable mode $\f_+(x)$, vanishes identically. We will not give a general proof of this statement here,
but one can understand it as follows. Generically, a non-zero $C[\f_-(x)]$ can only arise if there are
intermediate terms between the two modes, $\f_-(x)$ and $\f_+(x)$, in the asymptotic expansion (\ref{asymp_exp}). This
can happen only if $\D_+-\D_->\D_-$. However, (\ref{mass_range}) implies that for $d>2$, $\D_+-\D_-\leq \D_-$. Therefore,
at least for $d>2$ and for a mass in the range (\ref{mass_range}), one has
$\hp\sub{\D_+}=-(\D_+-\D_-)\f_{+}(x)/l$ exactly, and hence, $\f_+(x)$ {\em is} well defined on the
boundary in this case.

\subsection{Solution of the boundary value problem}

The general solution of the boundary value problem with the boundary condition $\d J(\f_-,\hp\sub{\D_+})=0$
consists in finding the most general regular solution of the bulk equations of motion as a functional
of the arbitrary source $J(\f_-,\hp\sub{\D_+})\equiv J(x)$. This involves two steps:
\begin{itemize}

\item {\em Radial problem}

One solves the radial equation of motion exactly, imposing regularity in the interior.
The result of this calculation is that (i) $\hp\sub{\D_+}$ is determined as a {\em non-local} functional of
$\f_-$\footnote{This should in no way be confused with the boundary condition, which itself
imposes {\em another} - {\em algebraic} - relation between the modes.} and (ii) the full bulk solution, $\f(r,\f_-(x))$, is obtained
as a non-local functional of $\f_-(x)$.

\item {\em Transverse problem}

To complete the solution of the boundary value problem, one needs to determine $\f_-(x)$  as a functional of
the arbitrary source $J(x)$. Having determined the functional $\hp\sub{\D_+}[\f_-]$ by solving the radial problem,
this is achieved by solving the equation
\be\label{transverse}
\framebox[1.7in]{
\begin{minipage}{3.5in}
\begin{equation*}
J(\f_-,\hp\sub{\D_+}[\f_-])=J(x),
\end{equation*}
\end{minipage}
}
\ee
for $\f_-[J]$. For the boundary conditions in Table \ref{bcs}, (\ref{transverse}) reads
\bea
& \f_-=J(x), & \text{Dirichlet},\NO\\
& -\hp\sub{\D_+}[\f_-]=J(x), & \text{Neumann},\NO\\
& -\hp\sub{\D_+}[\f_-]-f'(\f_-)=J(x), & \text{Mixed}.
\eea
Hence, the transverse problem is trivial for Dirichlet boundary conditions, but non-trivial for
Neumann and Mixed boundary conditions. In all cases, inserting the resulting solution $\f_-[J]$ back in
the bulk solution of the radial problem we obtain the full solution $\f(r,\f_-[J(x)])$.

\end{itemize}

Although the general prescription for solving the boundary value problem involves only these two simple steps,
in practice there are very few cases where one is able to carry out either of these two steps. In particular,
the bulk equations of motion are generically non-linear, which makes the solution of the radial problem
very difficult. On the other hand, if the function $f'(\f_-)$ is non-linear, then the solution of the
transverse problem (\ref{transverse}) becomes very difficult too. In the next section, however, we will
discuss a toy model for which it is possible to carry out the above prescription explicitly.

\subsection{The on-shell action and the AdS/CFT dictionary}

Assuming we have solved the boundary value problem with arbitrary sources to obtain the exact
solution $\f(r,\f_-[J])$, we can evaluate the on-shell action, $I[J]$. This involves three pieces:
the bulk action, $S$, the covariant boundary counterterms, $S_{\rm ct}$, and the boundary term, $S_{J}$, defining the boundary condition. Namely,
\be
I[J]=\left.\left(S+S_{\rm ct}+S_{J}\right)\right|_{\f},
\ee
where the limit $r\to\infty$ is implicit. By construction, the value of the sum of these terms
remains finite in this limit, and naturally, it is a functional of the source, $J$. The AdS/CFT dictionary,
or more generally the gravity/quantum field theory dictionary, identifies the on-shell action, $I[J]$, with
the generating functional of connected correlation functions of the operator dual to the scalar field $\f$.
Namely,
\be
Z[J]\equiv e^{-W[J]}=
\left\langle e^{-\int J\co}\right\rangle\approx e^{-I[J]},
\ee
where the $\approx$ sign in the last equality means that the identification
is understood in some certain limit, e.g. in the large $N$ limit, such that
supergravity is a good approximation to the field theory dual.\footnote{We will not be
specific about this limit since it depends crucially on the particular AdS/CFT duality.
For example, in the most studied AdS$_5$/CFT$_4$ duality between $\cn=4$ super Yang-Mills
and Type II B string theory, the supergravity approximation involves not only the large $N$
limit, but also the large 't Hooft coupling limit. However, in the duality between M-theory on
AdS$_4\times S^7$ and the $\cn=8$ SCFT in three dimensions, the supergravity approximation involves
only the large $N$ limit as there is no other free parameter in this case.} Since
the on-shell action, $I[J]$, is identified with the generating functional
of connected correlation functions, $W[J]$, its Legendre transform, $\G[\s]$,
given by
\be
\G[\s]=W[J]-\int_{\pa\cm}d^dx\sqrt{g\sub{0}}J\s,
\ee
is the effective action of the dual operator, i.e. the generating functional
of 1PI diagrams. In particular, the transverse problem (\ref{transverse})
has a direct interpretation in the dual field theory as the `gap equation' (\ref{gap}).
From Table \ref{dictionary} it is evident that, although the solution of the
transverse problem is required in order to evaluate the generating functional, $W[J]$,
for Neumann and Mixed boundary conditions (we have seen that the transverse problem is always trivial for
Dirichlet boundary conditions), only the solution of the {\em radial} problem is necessary to evaluate
the effective action, $\G[\s]$.
\begin{table}
\begin{center}
\begin{tabular}{|c|c|c|c|}
\hline &&&\\
 & Dirichlet & Neumann & Mixed  \\&&&\\
\hline\hline &&&\\
$J$ & $J_+\equiv\f_-$ & $J_-\equiv -\hp\sub{\D_+}$ &
      $J_{f_-}\equiv -\hp\sub{\D_+}-f'(\f_-)$\\ &&&\\
$\s$ & $\hp\sub{\D_+}$ & $\f_-$ &  $\f_-$ \\ &&&\\
$W[J]$ & $I_+[J_+]$ & $I_-[J_-]$ & $I_{f_-}[J_{f_-}]$\\ &&&\\
$\G[\s]$ &  $I_-[-\hp\sub{\D_+}]$ &
     $I_+[\f_-]$ &
     $I_+[\f_-]+\int_{\pa\cm} d^dx\sqrt{g\sub{0}} f(\f_-)$ \\ &&&\\
$\langle T^{ij}\rangle = \frac{-2}{\sqrt{g\sub{0}}}\frac{\d W}{\d g\sub{0}_{ij}}$
 & $-2\hp\sub{d}^{ij}$ & $-2\hp\sub{d}^{ij}-\f_- J_-g\sub{0}^{ij}$ &
    $-2\hp\sub{d}^{ij}-\left(f(\f_-)+\f_- J_{f_-}\right)g\sub{0}^{ij}$\\ &&&\\
    $\langle T^i_i\rangle $
 & $-(d-\D_+)J\s$ & $-(d-\D_-)J\s$ &
    $-(d-\D_-)J\s-d\left(f(\s)-\frac{\D_-}{d}\s f'(\s)\right)$\\ &&&\\
\hline
\end{tabular}
\end{center}
\caption{The gravity/QFT dictionary. }
\label{dictionary}
\end{table}

In Table \ref{dictionary} we summarize the identifications between the bulk and boundary
quantities, according to the gravity/quantum field theory dictionary, for the three
boundary conditions in Table \ref{bcs}. Note that, since the on-shell action for
Neumann boundary conditions is the Legendre transform of the on-shell action
for Dirichlet boundary conditions (see the boundary term $S_-$ in Table \ref{bcs}),
the effective action for Dirichlet boundary conditions is given by the on-shell
action for Neumann boundary conditions and vice versa. Moreover, the effective action
for Mixed boundary conditions is given by the on-shell action for Dirichlet boundary conditions
plus a term involving the function $f(\f_-)$. Comparing the effective actions for Neumann and Mixed boundary
conditions in Table \ref{dictionary} with the expressions for the undeformed and deformed effective
actions given in Table \ref{multi-trace-defs}, we reach the conclusion that
{\em the Mixed boundary conditions correspond to a multi-trace deformation of the QFT dual to the Neumann
boundary conditions} \cite{Witten:2001ua}.

In the penultimate row of Table \ref{dictionary} we show the renormalized VEV of the stress tensor of
the dual theory in terms of the renormalized radial momentum, $\hp\sub{d}^{ij}$, conjugate to the induced metric,
$\g_{ij}$, (see Appendix \ref{appendix}) for the three different boundary conditions. Note that the difference in
these expressions for the VEV of the stress tensor is due to the boundary terms in Table \ref{bcs}, which are
required to impose each boundary condition. Using the fact that the bulk equations of motion determine that the
trace of $\hp\sub{d}^{ij}$ is related to the renormalized scalar momentum, $\hp\sub{\D_+}$, by
$\hp\sub{d}=\D_-\hp\sub{\D_+}\f_-/2$ \cite{PS1}, these expressions allow us to write down the Conformal Ward
identity for each boundary condition. These Ward identities are shown in the last row of Table \ref{dictionary}.
We conclude that the Dirichlet and Neumann boundary conditions lead to a {\em conformal}
field theory dual, since $\langle T^i_i\rangle=0$ for vanishing source, while the Mixed boundary condition leads
to a conformal dual theory only if $f(\f_-)\propto \f_-^{d/\D_-}$. In that case, the Mixed boundary conditions
describe a {\em marginal} multi-trace deformation of the CFT dual to the Neumann boundary conditions. Moreover,
in the cases of a CFT dual, we see that the conformal dimension of the operator dual to the scalar field is $\D_+$
for Dirichlet boundary conditions and $\D_-$ for Neumann and Mixed boundary conditions. This is as expected, since
the leading asymptotic behavior of the physical solutions is determined by $\D_+$ for Dirichlet boundary conditions
(i.e. $\f_-=0$), but by $\D_-$ for Neumann and Mixed boundary conditions ($\f_-\neq 0$).

\section{Toy model}
\label{toy}
\setcounter{equation}{0}

The boundary value problem can be solved in complete generality, following the prescription outlined above, for a free
scalar field in a fixed AdS background with the action
\be
S=\int d^{d+1}x\sqrt{g}\left(\frac12g^{\m\n}\pa_\m\f\pa_\n\f+\frac12m^2\f^2
\right).
\ee
The metric here is (\ref{gf-metric}) with $\g_{ij}=e^{2r/l}\d_{ij}$, which is the metric of exact AdS$_{d+1}$
(more precisely $\mathbb{H}_{d+1}$) in the upper half plane coordinates. The equation of motion is of course the
Klein-Gordon equation (\ref{K-G}). Even though the fact that the bulk equation of motion is linear means that it
is possible to solve the radial problem exactly, the transverse problem remains in general intractable, except
for certain linear boundary conditions.

\subsection{General solution with linear boundary conditions}

\begin{flushleft}
{\bf Counterterms}
\end{flushleft}

In order to compute the renormalized momentum, as well as the on-shell action,
we need to know the covariant boundary counterterms. This is done, as we discussed above, by inserting the covariant
expansion (\ref{mom-exp}) of the canonical momentum, and of the radial derivative
(\ref{rad-der-exp}), into the equation of motion (\ref{K-G}). This iteratively determines \cite{Papadimitriou:2003is,thesis}
\bea\label{mom-counterterms}
\p\sub{\D_-}=-\frac{\D_-}{l}\f,\quad \p\sub{\D_-+2}=\frac{-l\square_\g\f}
{(d-2\D_--2)},\quad \p\sub{\D_-+4}=-\frac{l^3\left(-\square_\g\right)^2\f}
{(d-2\D_--2)^2(d-2\D_--4)},\quad \cdots
\eea
This procedure breaks down at order $\D_+$ leaving $\p\sub{\D_+}$ undetermined.
From (\ref{def-counterterms}) now we see that the counterterms are obtained by
integrating these momenta with respect to the scalar field. This gives \cite{lectures,Papadimitriou:2003is,thesis}
\be\label{toy-counterterms}
S_{\rm ct}=\frac12\int_{\S_r}d^dx\sqrt{\g}\f\left(-\frac{\D_-}{l}\f+
\frac{-l\square_\g\f}{(d-2\D_--2)}-\frac{l^3\left(-\square_\g\right)^2\f}
{(d-2\D_--2)^2(d-2\D_--4)}+\cdots\right).
\ee
If we restrict to the mass range (\ref{mass_range}), which is necessary in order to consider Neumann and Mixed
boundary conditions, then only the first term in (\ref{toy-counterterms}) is relevant since $\D_+\leq\D_-+2$ in
this case. For Dirichlet boundary conditions, however, more terms must be kept in general.

\begin{flushleft}
{\bf Radial problem}
\end{flushleft}

The most general solution of the radial part of the equation of motion (\ref{K-G}) that is regular
in the interior can be written in two equivalent forms. The first is an exact expression
for the canonical momentum as a covariant functional of the induced field $\f$, namely \cite{Papadimitriou:2003is,thesis}
\be\label{exact-mom}
\p_\f[\f]=\sqrt{\g}\dot{\f}=\sqrt{\g}\frac1l\left(-\frac d2-l\sqrt{-\square_\g}
\frac{K'_\n\left(l\sqrt{-\square_\g}\right)}
{K_\n\left(l\sqrt{-\square_\g}\right)}\right)\f,
\ee
where $\n=(\D_+-\D_-)/2$ and $K_\n$ is the modified Bessel function that is regular for large argument. Fourier
transforming (\ref{exact-mom}) and solving the resulting linear first order equation gives the second form
\be\label{exact-solution}
\framebox[4.3in]{
\begin{minipage}{4.7in}
\begin{equation*}
\f(r,\f_-(x))=\frac{l^\n}{2^{\n-1}\G(\n)}e^{-dr/2l}\int\frac{d^dp}{(2\p)^d}
\tilde{\f}_-(p)p^\n K_\n\left(lpe^{-r/l}\right)
e^{ip\cdot x},
\end{equation*}
\end{minipage}
}
\ee
where $\tilde{\f}_-(p)$, which appears as the integration constant of the first order equation,
is the Fourier transform of an arbitrary function $\f_-(x)$. Using the asymptotic form of the Bessel function,
$K_\n(lpe^{-r/l})\sim 2^{\n-1}l^{-\n}\G(\n)e^{\n r/l}p^{-\n}$, as $r\to\infty$, we see that $\f\sim e^{-\D_-r/l}\f_-(x)$
asymptotically, in agreement with (\ref{asymp_exp}).

The form (\ref{exact-mom}) is particularly useful because, by expanding the Bessel function for small argument,
one automatically obtains the covariant expansion (\ref{mom-exp}), but now including the renormalized momentum
$\p\sub{\D_+}$. For $0<\n<1$, which corresponds to the mass range\footnote{The endpoints, $\n=0,1$, correspond
respectively to the cases where the BF bound and the unitarity bound are saturated. Whenever $\n$ is an integer
the Bessel function involves logarithms. These cases can be treated similarly. See e.g.
\cite{lectures,Papadimitriou:2003is,thesis} for the case of Dirichlet boundary conditions, where logarithms also appear.}
(\ref{mass_range}), this is
\be\label{ren-mom}
\framebox[2.8in]{
\begin{minipage}{3.5in}
\begin{equation*}
\hp\sub{\D_+}[\f_-]=\left(\frac l2\right)^{2\n-1}\frac{\G(1-\n)}{\G(\n)}
\left(-\square\right)^\n\f_-.
\end{equation*}
\end{minipage}
}
\ee
This non-local relation is the essential ingredient in order to address the transverse problem.

\begin{flushleft}
{\bf Transverse problem}
\end{flushleft}

Since the transverse problem (\ref{transverse}) is trivial for Dirichlet boundary conditions, the solution of
the radial problem, (\ref{exact-mom}) or (\ref{exact-solution}), is sufficient to evaluate the on-shell action.
The result is shown in the second column of Table \ref{linear}. From Table \ref{dictionary} follows that this
also allows us to evaluate the {\em effective action} for Neumann and Mixed boundary conditions, since it is
directly related to the on-shell action for {\em Dirichlet} boundary conditions. For the Mixed boundary conditions
then, which include the Neumann as a special case, the effective action is
\be\label{eff-action}
\G_{f_-}[\f_-]=\int d^dx\left(\frac12\left(\frac l2\right)^{2\n-1}
\frac{\G(1-\n)}{\G(\n)}\f_-\left(-\square\right)^\n\f_-+f(\f_-)\right).
\ee
Functionally differentiating this with respect to the VEV, we explicitly see that the
transverse problem (\ref{transverse}), which, using (\ref{ren-mom}), takes the form
\be\label{b_eq}
-\left(\frac l2\right)^{2\n-1}\frac{\G(1-\n)}{\G(\n)}\left(-\square\right)^\n
\f_--f'(\f_-)=J(x),
\ee
is nothing but the gap equation (\ref{gap}).

Solving (\ref{b_eq}) for $\f_-[J]$ is in general not possible unless $f(\f_-)=\x \f_-^2$,
for some constant $\x$. As we have seen, this corresponds to a {\em double-trace} deformation
of the dual theory. With this special choice of Mixed boundary conditions, the transverse problem is easily
solved and the on-shell action can be evaluated. The results are shown in Table \ref{linear}. Moreover,
inserting the expressions for $\f_-[J]$ given in Table \ref{linear} in the solution (\ref{exact-solution}),
we obtain the full solution to the boundary value problem with the corresponding linear boundary conditions.
Note that the expression for the on-shell action in the presence of a double-trace deformation shown in
Table \ref{linear} is in complete agreement with the corresponding expressions in e.g. \cite{Hartman:2006dy},
but ours are manifestly cut-off independent.
\begin{table}
\begin{center}
\begin{tabular}{|c|c|c|c|}
\hline &&&\\
 & Dirichlet & Neumann & Mixed \\  &&&\\
\hline\hline &&&\\
$J$ & $J_+\equiv\f_-$  & $J_-\equiv -\hp\sub{\D_+}$ &  $J_{f_-}\equiv -\hp\sub{\D_+}-2\x\f_-$\\  &&&\\
$\f_-[J]$ & $J_+$ &$-2^{2\n-1}\frac{l\G(\n)}{\G(1-\n)}(-l^2\square)^{-\n}J_-$
& $-\left(2\x+\left(\frac 12\right)^{2\n-1}\frac{\G(1-\n)}{l\G(\n)}(-l^2\square)^\n\right)^{-1}J_{f_-}$\\  &&&\\
$I[J]$ &  $\frac{\G(1-\n)}{l2^{2\n}\G(\n)}\int J_{+} (-l^2\square)^{\n}J_+$
& $-\frac{l2^{2\n}\G(\n)}{4\G(1-\n)}\int J_-(-l^2\square)^{-\n}J_-$
& $-\frac14\int J_{f_-}\left(\x+\frac{\G(1-\n)}{l2^{2\n}\G(\n)}(-l^2\square)^{\n}\right)^{-1}J_{f_-}$\\ &&&\\
\hline
\end{tabular}
\end{center}
\caption{The solutions of the transverse problem and the on-shell action for all three linear
boundary conditions. $\x$ is an arbitrary constant corresponding to the deformation parameter
of the double-trace deformation.}
\label{linear}
\end{table}

\subsection{Vacua with non-linear boundary conditions}

Even though one cannot solve equation (\ref{b_eq}) in general for non-linear boundary
conditions, we can still find exact solutions of the corresponding sourceless equation.
The classification of such `vacuum'\footnote{By the term `vacuum' we do not necessarily refer
to a constant or time independent solution $\f_-$. Any solution to the sourceless equation
(\ref{transverse}) will be called a vacuum.} solutions is essential before one can solve
(\ref{b_eq}) perturbatively in the source, $J(x)$, around each vacuum. We will not
attempt a classification of the vacuum solutions for various choices of $f(\f_-)$ here. Instead,
we now give two examples of non-trivial vacua which are closely related to the
vacua we will construct later on for more realistic interacting theories.

\begin{flushleft}
{\bf Constant VEV}
\end{flushleft}

For any choice of the function $f(\f_-)$, a constant, $\f_-^*$, that extremizes $f(\f_-)$, i.e. $f'(\f_-^*)=0$,
is a solution of the sourceless equation (\ref{b_eq}). Indeed, from (\ref{eff-action}) we know that $f_-(\f)$ is
the effective potential of the dual theory. The Fourier transform of a constant $\f_-^*$ is a delta function
in momentum space, $\tilde{\f}_-^*=\f_-^*(2\p)^d\d^{(d)}(p)$. Inserting this into the general solution
(\ref{exact-solution}) we obtain an exact solution of the bulk equation of motion satisfying Mixed boundary conditions. Namely,
\be
\f(r,\f_-^*)=\frac{l^\n}{2^{\n-1}\G(\n)}e^{-dr/2l}\f_-^*\lim_{p\to 0}
p^\n K_\n\left(lpe^{-r/l}\right)e^{ip\cdot x}=e^{-\D_-r/l}\f_-^*.
\ee

\begin{flushleft}
{\bf Instantons}
\end{flushleft}

Non-constant solutions can also be found, at least for certain choices of the potential $f(\f_-)$.
To look for non-constant solutions, however, we need an explicit representation of the operator
$(-\square)^\n$. It is in fact easier to find a representation of the
inverse of this operator, which has the integral representation
\be
(-\square)^{-\n}\F(x)=\frac{\G\left(\frac d2-\n\right)}{2^{2\n}
\p^{d/2}\G(\n)}\int d^dy\frac{\F(y)}{|x-y|^{d-2\n}},\quad \n>0.
\ee
The sourceless equation (\ref{b_eq}) can then be written as the integral equation
\be\label{integral-eq}
\f_-(x)+\frac{\G\left(\frac d2-\n\right)}{2\p^{d/2}l^{2\n-1}\G(1-\n)}
\int d^dy\frac{f'(\f_-(y))}{|x-y|^{d-2\n}}=0.
\ee

We now look for solutions of the form
\be\label{instanton}
\f_-(x)=\frac{b}{|x|^c},
\ee
for a potential of the form $f(\f_-)=\x \f_-^\om$, where $b,c,\x,\om$ are constants.
Inserting these into (\ref{integral-eq}) and Fourier transforming determines that such a solution exists
only if $\om>2$, i.e. provided the boundary condition is {\em non-linear}, and also only if $\x<0$, i.e. when
the effective potential is {\em unbounded} from below. Moreover, $c$ and $b$ are given by
\be
c=\frac{2\n}{\om-2},\quad
b^{\om-2}=\frac{2 l^{2\n-1}\G(1-\n)\G\left(\frac d2-\frac{\n}
{\om-2}\right)\G\left(\frac{(\om-1)\n}{\om-2}\right)}
{\om|\x|\G(\n)\G\left(\frac{\n}{\om-2}\right)\G\left(\frac d2-\frac{(\om-1)\n}
{\om-2}\right)}.
\ee
Inserting (\ref{instanton}) into (\ref{exact-solution}) we obtain the exact bulk solution
\be\label{inst-sol}
\f(r,\f_-(x))=bl^{-\frac{2\n}{\om-2}}\frac{\G\left(\frac d2-\frac{\n}{\om-2}\right)
\G\left(\frac{(\om-1)\n}{\om-2}\right)}{\G\left(\frac d2\right)\G(\n)}
e^{-\left(\frac d2-\frac{\om\n}{\om-2}\right)\frac rl}
F\left(\frac{\n}{\om-2},\frac{(\om-1)\n}{\om-2};\frac d2;-\frac{1}{l^2}
e^{2r/l}x^2\right).
\ee
The asymptotic form of this solution is
\be
\f(r,\f_-(x))\sim \left(\frac{b}{|x|^c}\right)e^{-\D_-r/l}+\frac{l}{2\n}\om\x
\left(\frac{b}{|x|^c}\right)^{\om-1}e^{-\D_+r/l},
\ee
i.e. $\f_+=\frac{l}{2\n}\om\x\f_-^{\om-1}$. Since $\hp\sub{\D_+}=-2\n\f_{+}(x)/l$, it follows
that (\ref{inst-sol}) indeed satisfies Mixed boundary conditions with $f(\f_-)=\x\f_-^\om$, for
$J_{f_-}=-\hp\sub{\D_+}-f'(\f_-)=0$.

This Euclidean solution is in fact analogous to the instanton solution found in \cite{dHPP} for
a scalar field conformally coupled to four-dimensional gravity, which we will revisit and generalize
below. As for the instanton solution of \cite{dHPP}, (\ref{inst-sol}) exists only when the effective
potential is unbounded from below, i.e $\x<0$, which means that the deformation induces an instability
in the boundary CFT. In particular,  (\ref{inst-sol}) describes the decay process of the
trivial vacuum at $\f_-=0$ to an instability region at $\f_-\to\infty$. The decay rate is given by
\cite{Coleman:1980aw}
\be
\cp\propto e^{-\left.\G_f\right|_{\rm inst.}},
\ee
where the value of the effective action (\ref{eff-action}) evaluated on the instanton solution is
\be
\left.\G_{f}\right|_{\rm inst.}=\frac{d(\om-2)|\x|}{2\left(d-\frac{2\n\om}
{\om-2}\right)}b^{\om}{\rm Vol}(S^d).
\ee

\section{Effective action from Hamilton's characteristic function}
\setcounter{equation}{0}
\label{method}

The free scalar field in a fixed AdS background is a useful example as a boundary value problem
in AlAdS spaces, but, in the context of the AdS/CFT correspondence, it can only give information on
the dual CFT at the conformal vacuum. As soon as the scalar field acquires a non-zero VEV conformal
invariance is broken and one must couple the scalar to dynamical gravity in order to study holographically
the dual field theory. In particular, although the conformal vacuum generically remains a vacuum of the
dual theory when the latter is deformed by a multi-trace deformation, the deformation may not only
destabilize the conformal vacuum, but also it will generically introduce new vacua. AdS/CFT relates the
problem of stability of the conformal vacuum under multi-trace deformations to the stability of AdS under
the corresponding boundary conditions
on the dual bulk fields.

Both the non-perturbative stability of the conformal vacuum and the possible appearance of new
vacua due to a generic multi-trace deformation can be addressed if one knows
the effective action for the deforming operator in the dual field theory. In the large $N$
limit, the AdS/CFT dictionary relates the effective action of the boundary theory to the
on-shell supergravity action.  More specifically, the on-shell action with Dirichlet boundary
conditions is related to the effective action of the theory with Neumann or Mixed boundary conditions
and vice versa (see Table \ref{dictionary}). However, computing the on-shell supergravity action
{\em non-perturbatively} in the scalar field - which is required if non-perturbative stability is
to be addressed - is of course not an easy task. Even though the boundary metric $g\sub{0}_{ij}$
can be set to a fixed value for this computation, since we are only interested in the back-reaction
of the scalar field on the bulk metric, the bulk equations remain highly non-linear and generically
too difficult to solve. Nevertheless, there is a systematic way to approximate the effective action
away from the conformal vacuum. Since, for conformal boundary conditions, conformal invariance
is only broken spontaneously by the non-zero VEV of the scalar field, in a vacuum of non-zero
scalar VEV the two-point function of the scalar operator always contains a massless pole,
corresponding to the Goldstone boson of spontaneously broken scale invariance, which dominates
the two-point function for small momenta. This massless pole gives, via the Legendre transform, a standard
quadratic kinetic term in the effective action for the VEV of the scalar operator. It follows that,
at least for conformal boundary conditions, the effective action admits a derivative expansion away from
the conformal vacuum. But since we know that a generic multi-trace deformation simply modifies
the effective potential in the large $N$ limit, the above argument implies that the effective
action always admits a derivative expansion away from the conformal vacuum. As we will now see, this fact
allows us to systematically construct the effective action to any order in derivatives, although
the computation quickly becomes tedious.

According to standard practice, to evaluate the renormalized on-shell action one needs to compute two
related - yet distinct - quantities. Namely, the covariant boundary counterterms, $S_{\rm ct}$, and the
regularized action, $I_r$ (see (\ref{reg-action})). Let us take the opportunity here to emphasize that
the split of the computation into two separate computations of $S_{\rm ct}$ and $I_r$ is only an artificial
split reflecting the difference in technical difficulty in computing these two quantities.
While the counterterms can always be computed in full generality by some version of
holographic renormalization, the computation of $I_r$ is far more difficult and usually
requires some approximation. However, if one were able to compute $I_r$ exactly, then
the counterterms, and hence the renormalized on-shell action, can be immediately deduced by
expanding $I_r$ in eigenfunctions of the dilatation operator. In fact, we saw such an
example in Section \ref{toy}, where the exact expression for the momentum (\ref{exact-mom})
enabled us to simply read off the renormalized momentum (\ref{ren-mom}). But the advantage of the
expansion in eigenfunctions of the dilatation operator is that it works equally well
even when $I_r$ is only known in some approximation, be that a small source expansion or a
derivative expansion. In particular, instead of the traditional 3-step approach: 1. compute
$S_{\rm ct}$ in full generality, 2. compute $I_r$ in some approximation 3. reduce the
counterterms in the approximation used for evaluating $I_r$, we will use the more efficient
2-step approach: 1. compute $I_r$ in whatever approximation is suitable 2. extract the
renormalized part of $I_r$ by expanding it in eigenfunctions of the dilatation operator
and keeping terms of weight zero. This technique was applied to compute renormalized
correlation functions in \cite{PS2}, which involved evaluating $I_r$ in a small source
approximation. It can equally well be applied in the case of a derivative expansion of
$I_r$, which is the relevant approximation here. Hence, we simply need to worry about
the evaluation of the regularized on-shell action $I_r$, as the renormalized action
can effectively be read off $I_r$.

The most direct way to compute the regularized on-shell action is via the radial Hamilton-Jacobi formalism
\cite{dBVV}.\footnote{It is worth pointing out that the interpretation of the Hamilton-Jacobi equation for
the on-shell action as the Callan-Symanzik equation for the generating functional of the dual operator in
\cite{dBVV} has an obvious analogue when the AdS/CFT dictionary identifies the on-shell action
with the effective action of the dual operator.} This amounts to solving the two functional equations
resulting from inserting the radial momenta, given as functional derivatives of the regularized on-shell
action with respect to the induced fields on $\S_r$ as in (\ref{HJ}), in the Hamiltonian and momentum
constraints (\ref{constraints}). Although the resulting functional differential equations are generically
too complicated to solve, their virtue is that they directly determine the regularized on-shell action.
In practice, this approach is useful if one is able to write down the most general ansatz for $I_r$ in a
certain approximation, or if one is interested
in the `minisuperspace' of a certain class of solutions, since then the Hamilton-Jacobi equation
can be simplified drastically. Although the local part of $I_r$, which will be removed
by the boundary counterterms, always takes the form of a derivative expansion and so
can be determined by an obvious ansatz as in \cite{dBVV}, here we are interested in
the {\em non-local} part of the regularized action. Finding a suitable ansatz for this
non-local part is a much more difficult question. However, this crucially
depends on the physical interpretation of the regularized action. In particular,
while a local derivative expansion of the non-local part of $I_r$ is useless if $I_r$
is identified with the generating functional of the dual operator, since a derivative
expansion only gives contact terms in the corresponding correlation functions, it {\em does}
makes sense to expand the non-local part of $I_r$ in a derivative expansion if it
is interpreted as the effective action of the dual operator. Indeed, we have argued above
that the effective action admits such a local approximation away from the conformal vacuum.

Our method for evaluating the regularized on-shell action then consists of two complementary
computations, involving different but not mutually exclusive approximations. First,
we start from the undeformed CFT on a (nearly) flat boundary, in which case the regularized
action in the two-derivative approximation takes the form\footnote{See \cite{Parry:1993mw} for
a systematic approach to solving the Hamilton-Jacobi equation in a derivative (long wavelength)
expansion in a similar context.}
\be\label{action-ansatz}
I_r=\int_{\S_r}d^dx\sqrt{\g}\left(W(\f)+Z(\f)R[\g]+\frac12M(\f)\g^{ij}\pa_i\f\pa_j\f\right),
\ee
where $W(\f)$, $Z(\f)$ and $M(\f)$ are functions of the scalar field to be determined.
Differentiating (\ref{action-ansatz}) with respect to the induced metric and the scalar field yields
respectively the momenta (see (\ref{HJ}))
\bea\label{momenta-ansatz}
\p^{ij}&=&\sqrt{\g}\left\{\left(Z''-\frac12M\right)\pa^i\f\pa^j\f+Z'D^iD^j\f-ZR^{ij}\right.\NO\\
&&\left.+\frac12\g^{ij}\left[W+ZR-2Z'\square_\g\f+\left(\frac12M-2Z''\right)\pa^k\f\pa_k\f\right]\right\},\NO\\
\p_\f&=&\sqrt{\g}\left\{W'+Z'R-\frac12M'\pa^k\f\pa_k\f-M\square_\g\f\right\}.
\eea
These momenta automatically satisfy the momentum constraint in (\ref{constraints}), which is simply a consequence
of the invariance of (\ref{action-ansatz}) under $\S_r$ diffeomorphisms. Inserting the
momenta (\ref{momenta-ansatz}) into the Hamiltonian constraint in (\ref{constraints}) leads to a set of ordinary
differential equations for the functions $W(\f)$, $Z(\f)$ and $M(\f)$, which of course
depend on the form of the Hamiltonian. Having determined these functions, the regularized
action is expanded in eigenfunctions of the dilatation operator (which in this case amounts
to a simple Taylor expansion of $W(\f)$, $Z(\f)$ and $M(\f)$) and the term of zero dilatation
weight, corresponding to the renormalized action, is isolated. The structure of the resulting renormalized
two-derivative effective action is largely universal as it is determined by conformal invariance.
However, there is a free parameter which is left undetermined both by the above procedure
and by conformal invariance. This raises the question as to what is the significance of this
parameter and how it can be fixed. Another question that is left unanswered by the above computation
is how one can evaluate the effective action with general boundary conditions, i.e. the effective action of the
deformed CFT. The diffeomorphism invariance of the ansatz (\ref{action-ansatz}) implicitly assumes that conformally
invariant boundary conditions are imposed, which is reflected in the fact that the renormalized action one obtains
from this ansatz is conformally invariant. However, a generic boundary condition corresponding to a relevant
deformation will break this invariance. So, how then are relevant deformations accommodated in the Hamilton-Jacobi
formalism? The answer to both questions is clearly that the ansatz (\ref{action-ansatz}) is too restrictive.

Instead of looking for a suitable generalization of the ansatz (\ref{action-ansatz}), however, we will try to
solve the Hamilton-Jacobi equation exactly - without any ansatz - but in a zero-derivative approximation.
That is, assuming the metric and the scalar are functions of the radial coordinate only. Clearly, this determines the
most general effective potential and hence it must account for any multi-trace deformation. Moreover, this calculation
can in principle be done for any boundary, not just a (nearly) flat boundary as was assumed in (\ref{action-ansatz}).
The result will be the exact effective potential on the given boundary. As we will see, this answers the first question,
since if one is able to evaluate the effective potential on, say, the sphere, then expanding this for small curvature
and comparing with the result of the previous calculation based on the ansatz (\ref{action-ansatz}) fixes the
undetermined parameter in the two derivative effective action.

Interestingly, upon a choice of a boundary manifold, the zero-derivative approximation amounts to looking at the
`minisuperspace' of certain domain-wall like solutions. In particular, for a boundary that cannot be
written as the direct product of two sub-manifolds, this approximation corresponds to looking for domain walls of the form
\be\label{generic-dw}
ds^2=dr^2+e^{2A(r)}g\sub{0}_{ij}(x)dx^idx^j,\quad \f=\f(r),
\ee
where $g\sub{0}_{ij}(x)$ is a metric independent of the radial coordinate $r$. The equations of motion
require that $g\sub{0}_{ij}(x)$ is maximally symmetric, $R[g\sub{0}]_{ij}=\frac1dR[g\sub{0}]g\sub{0}_{ij}$,  and has
locally constant scalar curvature, $R[g\sub{0}]=kd(d-1)/l^2$, where $k=0,\pm 1$. The hypersurface
$\S_r$ then can be $S^d$, $\mathbb{R}^d$ or $\mathbb{H}_d$, or a quotient of these by a discrete subgroup
of their isometry group. Replacing the induced metric $\g_{ij}$ as a dynamical field by the warp factor, $A(r)$, and
the canonical momentum $\p^{ij}$ by the momentum, $\p_A$, conjugate to $A$, which is defined via $\p^{ij}\d\g_{ij}=\p_A\d A$,
reduces the Hamilton-Jacobi equation to a PDE for the effective potential as a function of the two variables $A$ and $\f$.
This can then be viewed as the Hamilton-Jacobi equation
for a standard classical mechanics problem for the generalized coordinates $A$ and $\f$. As is well known
from Hamilton-Jacobi theory \cite{Goldstein}, the general solution of the equations of motion, i.e.
the most general solution of the form (\ref{generic-dw}) in this case, can be obtained from any complete
integral of the Hamilton-Jacobi equation, which in this case contains one arbitrary integration
constant.\footnote{This is because there is no explicit dependence on the radial `time'. The
characteristic function for $n$ variables contains $n-1$ arbitrary constants \cite{Goldstein}.}
However, the Hamilton-Jacobi equation admits more than one complete integrals. In fact, the general
solution of the Hamilton-Jacobi equation contains an arbitrary function - not just a constant.
As we will show below, this freedom in choosing a complete integral for the Hamilton-Jacobi equation
in the zero-derivative approximation corresponds precisely to the choice of boundary conditions.

Put together then, the two-derivative effective action with conformal boundary conditions based on
the ansatz (\ref{action-ansatz}), and the `minisuperspace' approximation of the Hamilton-Jacobi
equation to solutions of the form (\ref{generic-dw}), completely determine the two-derivative effective
action of the dual operator on $S^d$, $\mathbb{R}^d$ or $\mathbb{H}_d$ (or any of their quotients)
and for {\em any} boundary conditions. Moreover, this computation can be generalized to other boundaries
too. For example, one can compute the two-derivative effective action on $\mathbb{R}\times S^{d-1}$
by solving the Hamilton-Jacobi equation for metrics of the form
\be
ds^2=dr^2+e^{2A(r)}d\t^2+e^{2B(r)}d\Om_{d-1}^2,
\ee
instead of the domain walls (\ref{generic-dw}).

\section{Minimal coupling}
\label{minimal}
\setcounter{equation}{0}

In this section we apply the above method to the system of a single scalar field minimally coupled to
gravity, which is described by the action
\be\label{bulk_minimal_action}
S=\int_{\cm}d^{d+1}\sqrt{g}\left(-\frac{1}{2\k^2}R+\frac12
g^{\m\n}\pa_\m\f\pa_\n\f+V(\f)\right),
\ee
where $\k^2=8\p G_{d+1}$ is the gravitational constant.\footnote{See Appendix \ref{appendix} for a
detailed discussion of the variational problem and a derivation of the appropriate boundary terms.}
An action of this form generically arises as a consistent truncation of some gauged
supergravity\footnote{Of course, the `consistency' of the truncation must be checked at the level
of the equations of motion.}, but we need not be specific about the embedding of (\ref{bulk_minimal_action})
into a particular gauged supergravity at this point. We will later give an example of a potential
which allows this action to be embedded into $\cn=8$ gauged supergravity in four dimensions and, hence,
be uplifted to 11-dimensional supergravity, but we would like to keep the discussion as general as
possible here.

The action (\ref{bulk_minimal_action}) possesses an AdS vacuum of radius $l$ provided the potential has
a negative extremum at $\f=\f_o$, such that $V'(\f_o)=0$, $V(\f_o)=-d(d-1)/2\k^2l^2$. It follows that,
in the vicinity of this extremum, the potential takes the form
\be\label{potential_asymp}
V(\f)=-\frac{d(d-1)}{2\k^2l^2}+\frac12 m^2(\f-\f_o)^2+o((\f-\f_o)^2),
\ee
where $m$ is the mass of the scalar field. Note that, unless the potential is exactly constant, the
location of the extremum is at some fixed value, $\f_o$, which can be set to zero by a redefinition of
the scalar field. Moreover, if the mass vanishes, then the potential must be constant, or else the equations
of motion eliminate the mode $\f_-(x)$ in the expansion (\ref{asymp_exp}). Below we will focus on
masses in the range $-(d/2)^2\leq m^2l^2<0$, excluding the case of a constant potential.

\subsection{Two-derivative effective action for conformal boundary conditions}

Our first task is to determine the renormalized on-shell action corresponding to
conformal boundary conditions using the ansatz (\ref{action-ansatz}). Using the Hamiltonian (\ref{Hamiltonian-min}),
which is relevant for the action (\ref{bulk_minimal_action}), the Hamiltonian constraint leads to three independent
equations for the functions $W(\f)$, $Z(\f)$ and $M(\f)$. Namely,
\bea
\label{superpotential}
&&V(\f)=\frac12\left(W'{}^2-\frac{d\k^2}{d-1}W^2\right),\\
\label{Z}
&&W'Z'-\k^2\frac{d-2}{d-1}WZ+\frac{1}{2\k^2}=0,\\
\label{M}
&&M=2\k^2\frac{W}{W'}Z'.
\eea
The last equation gives explicitly the function $M(\f)$ in terms of $W$ and $Z$. Moreover,
the second equation is a linear equation for $Z$, whose solution in terms of $W$ is
\be\label{Z-sol}
Z(\f)=-\frac{1}{2\k^2}Z_o\int^\f\frac{d\bar{\f}}{W'Z_o},\quad
Z_o(\f)\equiv\exp\left(\k^2\frac{d-2}{d-1}\int^\f d\bar{\f}\frac{W}{W'}\right).
\ee
The regularized two-derivative effective action is therefore determined by the non-linear equation
(\ref{superpotential}) for the function $W(\f)$, which we will call `fake superpotential' for reasons
that will become clear later on. Equation (\ref{superpotential}) can be transformed \cite{PS1} into the
standard form of Abel's equation of the first kind \cite{Kamke}. Although its general solution is not
known for an arbitrary scalar potential $V(\f)$, for certain choices of the potential it falls into some
of the known integrability classes of Abel's equation and it can be solved exactly. We will discuss such
an example below, but in order to determine the renormalized action we need not solve (\ref{superpotential})
exactly. Indeed, we will now show that some general features of the solutions of equation (\ref{superpotential})
are sufficient for this purpose.

Firstly, from equations (\ref{momenta-ansatz}) follows that the asymptotic form of the induced metric and of
the scalar field is determined by the form of $W(\f)$ in the vicinity of $\f=0$. In particular, requiring the
metric to be AlAdS, fixes $W(0)=-(d-1)/\k^2 l$. This, in combination with the form (\ref{potential_asymp})
of the scalar potential near $\f=0$, determines, depending on the value of the scalar mass, the allowed
behaviors of $W(\f)$ around $\f=0$, which are shown in Table \ref{W-asymptotics}.
\begin{table}
\begin{center}
\begin{tabular}{|c|l|}
\hline &\\
$m^2l^2$ & $W(\f)$ \\ &\\
\hline\hline &\\
$-(d/2)^2<m^2l^2<0$ & $W_\pm(\f)=-\frac{(d-1)}{\k^2l}-\frac{1}{2l}\D_\pm \f^2+o\left(\f^2\right)$ \\&
\\\hline &\\
\multirow{3}{*}{$m^2l^2=-(d/2)^2$} & $W_+(\f)=-\frac{(d-1)}{\k^2l}-\frac{1}{2l}\frac d2 \f^2+o\left(\f^2\right)$ \\ &\\
                  & $W_-(\f)= -\frac{(d-1)}{\k^2l}-\frac{1}{2l}\frac d2 \f^2\left(1+\frac{1}{\log\f}\right)
+o\left(\frac{\f^2}{\log\f}\right)$ \\& \\ \hline
\end{tabular}
\end{center}
\caption{The allowed behavior of the fake superpotential in the vicinity of the AdS critical point at
$\f=0$.}
\label{W-asymptotics}
\end{table}
Note that there are two possible asymptotic behaviors in each case. The $W_+$ solutions
imply that the non-normalizable mode, $\f_-(x)$, vanishes in the corresponding solution
of the bulk equations of motion, which is obtained from $W(\f)$ via (\ref{momenta-ansatz}).
On the other hand, the $W_-$ solutions allow for a non-zero $\f_-(x)$. Since $\D_\pm$ are
the two roots of the equation $m^2l^2=\D(\D-d)$, the requirement that $W(\f)$ and hence $\D_\pm$
are real imposes the well known BF bound $m^2l^2\geq -(d/2)^2$ \cite{Breitenlohner:1982bm, Townsend:1984iu}.
Further classification of the possible solutions of equation (\ref{superpotential}) is facilitated by
the following lemma.
\begin{lemma}\label{first-lemma}
Provided the BF bound holds and $\D_->0$, any $W_-$ solution of equation (\ref{superpotential}) lies
on a continuous family of $W_-$ solutions while any $W_+$ solution is isolated, or corresponds to an end
point of an one-parameter family of $W_-$ solutions, at an infinite distance in parameter space from any
given $W_-$ solution.
\end{lemma}
To prove this lemma, we will assume that the original solution $W(\f)$
lies on a one-parameter family of solutions. In the case of $W_+$ solutions
we will show that this always leads to a contradiction, while for $W_-$ solutions
we construct explicitly the one-parameter family of solutions in the
neighborhood of $W(\f)$ when $\D_->0$. Suppose then that the solution $W(\f)$ lies on a continuous family
of solutions parameterized by the integration constant $\x$, chosen such that $\x=0$ corresponds to $W(\f)$.
The one-parameter family of solutions around $W(\f)$ then takes the form
\be\label{def-exp}
W(\f;\x)=W(\f)+\x W^{(1)}(\f)+O(\x^2),
\ee
where
\be\label{deformation}
W^{(1)}(\f)=\exp\left(\frac{d\k^2}{d-1}\int^\f d\tilde{\f}
\frac{W(\tilde{\f})}{W'(\tilde{\f})}\right).
\ee
Let us now assume that the original solution $W(\f)$ is of $W_+$ type.
Using the asymptotic form of $W_+$ solutions for the various masses
given in Table \ref{W-asymptotics}, we can deduce the corresponding asymptotic behavior of
$W^{(1)}(\f)$. One finds,
\be\label{W+asymp}
W^{(1)}(\f)\sim \left\{\begin{matrix}
\f^{d/\D_+}, & -(d/2)^2<m^2l^2<0,\\
\f^2, & m^2l^2=-(d/2)^2.
\end{matrix}\right.
\ee
Since $1<d/\D_+ <2$ when $ -(d/2)^2<m^2l^2<0$, we see that in all cases,
if one starts with a $W_+$ solution, the deformed solution $W(\f;\x)$
has asymptotics which are not compatible with the asymptotics
in Table \ref{W-asymptotics}, which any solution must obey. We have therefore reached a contradiction
and we conclude that any $W_+$ type solution is isolated. On the other hand, if the original
solution is of $W_-$ type, then the deformation (\ref{deformation}) behaves asymptotically as
\be\label{W-asymp}
W^{(1)}(\f)\sim \left\{\begin{matrix}
\f^{d/\D_-}, & -(d/2)^2<m^2l^2<0,\\
\frac{\f^2}{(\log \f)^2}, & m^2l^2=-(d/2)^2.
\end{matrix}\right.
\ee
If $\D_->0$, the BF bound ensures that $d/\D_->2$ and so we see that in this case the asymptotic form
of the deformation $W^{(1)}(\f)$ is {\em subleading} relative to the asymptotic behavior  of the original
solution. Hence, the deformed solution does exist, at least in the neighborhood of the original solution,
and it is of $W_-$ type for any (finite) value of the deformation parameter.\footnote{Lemma \ref{second-lemma} below
guarantees that the asymptotics is not affected by the higher order in $\x$ terms either.}  This completes
the proof of the above lemma.
\begin{flushright}
$\square$
\end{flushright}
The last claim in the the proof of the above lemma, namely that, when the deformation exists, the deformed
solution (\ref{def-exp}) remains of $W_-$ type to all orders in the deformation parameter $\x$,
follows from the next lemma.
\begin{lemma}\label{second-lemma}
The deformation parameter $\x$ can be chosen such that all higher-than-first
order in $\x$ terms are also asymptotically subleading relative to
$W^{(1)}(\f)$, i.e. such that $W^{(n)}(\f)=o\left(W^{(1)}(\f)\right)$ as
$\f\to 0$ for all $n>1$.
\end{lemma}
To prove this statement, we expand the deformed solution $W(\f;\x)$ as
\be\label{def-exp-full}
W(\f;\x)=\sum_{n=0}^\infty\x^n W\super{n}(\f),
\ee
where $W\super{0}(\f)\equiv W(\f)$ denotes the undeformed solution. Inserting
this in (\ref{superpotential}) we determine that $W\super{1}(\f)$ is
given by (\ref{deformation}) while for $n>1$
\be\label{W-n}
W\super{n}(\f)=W\super{1}(\f)\int^\f d\tilde{\f}\frac{Q\super{n}}
{W\super{0}'W\super{1}},
\ee
where
\be
Q\super{n}\equiv-\frac12\sum_{m=1}^{n-1}\left(W\super{m}'W\super{n-m}'
-\frac{d\k^2}{d-1} W\super{m}W\super{n-m}\right).
\ee
Using the asymptotics for the $W_-$ type solutions given in Table \ref{W-asymptotics}, we can
show that there exists a unique value of the integration constant implicit in (\ref{W-n}), such that
$W\super{2}(\f)=o\left(W^{(1)}(\f)\right)$ as $\f\to 0$. Since the
integration constant in (\ref{W-n}) simply multiplies the homogeneous
solution $W\super{1}$, it follows that any other value of the integration
constant can be absorbed in the definition of the deformation parameter
$\x$. A simple inductive argument can now be used to complete the proof for any order $n>1$.
\begin{flushright}
$\square$
\end{flushright}
Note that in order to determine whether a given $W_+$ solution is the end point of an one-parameter family
of $W_-$ solutions we need to treat the deformation non-perturbatively, which is to say, we must be able to
solve (\ref{superpotential}) exactly. We will consider a case for which this is possible in the next section, where
we will show that indeed the $W_+$ solution is the end point of an one-parameter family of $W_-$ solutions.

The final ingredient we need to evaluate the renormalized on-shell action for both $W_+$ and $W_-$ type solutions
is the asymptotic form of the functions $Z(\f)$ and $M(\f)$, given in Table \ref{ZM-asymptotics}, which follows
from that of the fake superpotential in Table \ref{W-asymptotics} via equations (\ref{Z}) and (\ref{M}). With this
last piece of information then we now only need to insert the functions $W(\f)$, $Z(\f)$ and $M(\f)$ in the regularized
action (\ref{action-ansatz}) and identify the piece of zero dilatation weight. This is straightforward and results
in the renormalized actions given in Table \ref{ren-eff-actions}, but a couple of subtle, yet important, points are
worth mentioning. Firstly, there is the freedom of adding extra finite local counterterms to $S_{\rm ct}$, which we mentioned in Section \ref{bvp}.
In this case it is manifested by the arbitrariness of the parameter $\x$ in the effective actions obtained from $W_-$ solutions. Different $W_-$
solutions lead to a different value for this parameter and so a definite choice of counterterms amounts to setting
this parameter to zero for a {\em particular} $W_-$ solution. Once this choice is made, however, all other $W_-$ solutions
will necessarily have a {\em non-zero} $\x$. It is clear then that for Mixed boundary conditions the freedom of adding finite local counterterms simply
corresponds to the choice of what one defines to be the `undeformed' theory - and {\em not} to a renormalization scheme dependence as is the case for
Dirichlet boundary conditions.  
In writing the effective actions in Table \ref{ren-eff-actions} we have picked
a random $W_-$ solution and we have assigned it the value $\x=0$. We will see later that if the system (\ref{bulk_minimal_action})
is embedded in some gauged supergravity, a natural choice for the solution that defines the zero of the parameter $\x$ is
the true superpotential of the theory, provided, of course, it is a $W_-$ type solution of (\ref{superpotential}). A second
minor point to note is that for $d=2$ the leading term of $Z(\f)$ gives both a divergent term, which is removed by the
counterterms, as well as the finite piece that contributes to the renormalized action. This is clear if one splits the
logarithm as $\log\f\sim\log e^{-\D_\pm r/l}+\log\f_\pm(x)$. The first piece then gives the usual logarithmically
divergent term associated with the conformal anomaly \cite{HS}, while the second piece gives the finite contribution to
the renormalized action. Note also that we have added an arbitrary function $f(\f_-)$ in the effective actions arising
from $W_-$ solutions, which corresponds to a general multi-trace deformation. Although the above argument does not
account for these terms, we have already seen that they arise from a choice of boundary conditions and we will discuss
how they can be accommodated in the the Hamilton-Jacobi setting below. Finally, except from the parameter $\x$ that
appears only in the $W_-$ effective actions, and which as we just saw corresponds to a choice of boundary conditions,
the effective actions in Table \ref{ren-eff-actions} also depend on the undetermined parameters $c_\pm$. These
parameters multiply a conformally invariant combination of the two two-derivative terms (for $d>2$), and can be
determined as we will see by computing the effective potential on a curved boundary, which fixes the coefficient of the curvature term.
\begin{table}
\begin{center}
\begin{tabular}{|c|c|l|l|}
\hline &&&\\
&$m^2l^2$ & $Z(\f)$ & $M(\f)$ \\&&&\\
\hline\hline &&&\\
\multirow{7}{*}{$d>2$}&$-(d/2)^2<m^2l^2$ & $Z_\pm(\f)\sim -\frac{l}{2\k^2(d-2)}+c_\pm\f^{\frac{d-2}{\D_\pm}}$ &
$M_\pm(\f)\sim c_\pm\frac{2(d-1)(d-2)}{\D_\pm^2}\f^{\frac{d-2}{\D_\pm}-2}$\\&&& \\ \cline{2-4}  &&&\\
&\multirow{3}{*}{$-(d/2)^2$} & $Z_+(\f)\sim -\frac{l}{2\k^2(d-2)}+c_+\f^{\frac{2(d-2)}{d}}$ & $M_+(\f)\sim c_+\frac{8(d-1)(d-2)}{d^2}
\f^{-\frac{4}{d}}$\\ &&&\\
         &  &       $Z_-(\f)\sim -\frac{l}{2\k^2(d-2)}+c_-\left(\frac{\f}{\log\f}\right)^{\frac{2(d-2)}{d}}$
&$M_-(\f)\sim c_-\frac{8(d-1)(d-2)}{d^2}
\frac{1}{\f^2}\left(\frac{\f}{\log\f}\right)^{\frac{2(d-2)}{d}}$\\ &&&\\ \hline &&&\\
\multirow{1}{*}{$d=2$} & $m^2<0$ & $Z_\pm(\f)\sim \frac{l}{2\k^2\D_\pm}\log\f+c_\pm$ & $M_\pm(\f)\sim\frac{l}{\k^2\D_\pm^2}\frac{1}{\f^2}$\\
 &&&\\ \hline
\end{tabular}
\end{center}
\caption{The asymptotic behavior of the functions $Z(\f)$ and $M(\f)$ following from that of the fake superpotential, $W(\f)$,
in Table \ref{W-asymptotics} via equations (\ref{Z}) and (\ref{M}). $c_\pm$ are arbitrary constants corresponding to the
integration constant of equation (\ref{Z}). }
\label{ZM-asymptotics}
\end{table}
\begin{table}
\begin{center}
\begin{tabular}{|c|c|c|}
\hline&&\\
$d$ & $W$ & $\G[\s]$ \\&&\\
\hline\hline &&\\
\multirow{4}{*}{$>2$}&$+$ & $c_+\int_{\pa\cm}d^dx\sqrt{g\sub{0}}\left(\f_+^{\frac{d-2}{\D_+}}R[g\sub{0}]
+\frac{(d-1)(d-2)}{\D_+^2}\f_+^{\frac{d-2}{\D_+}-2}g\sub{0}^{ij}\pa_i\f_+\pa_j\f_+\right)$ \\&&\\
& $-$ & $\int_{\pa\cm}d^dx\sqrt{g\sub{0}}\left(\x\f_-^{\frac{d}{\D_-}}+f(\f_-)+c_-\f_-^{\frac{d-2}{\D_-}}R[g\sub{0}]
+c_-\frac{(d-1)(d-2)}{\D_-^2}\f_-^{\frac{d-2}{\D_-}-2}g\sub{0}^{ij}\pa_i\f_-\pa_j\f_-\right)$ \\&&\\\hline&&\\
\multirow{4}{*}{$2$} & $+$ & $\frac{l}{2\k^2\D_+}\int_{\pa\cm}d^dx\sqrt{g\sub{0}}\left(\log\f_+R[g\sub{0}]
+\frac{2}{\D_+}\f_+^{-2}g\sub{0}^{ij}\pa_i\f_+\pa_j\f_+\right)+c_+\chi$ \\ &&\\
& $-$ & $\int_{\pa\cm}d^2x\sqrt{g\sub{0}}\left(\x\f_-^{\frac{2}{\D_-}}+f(\f_-)
+\frac{l}{2\k^2\D_-}\left(\log\f_-R[g\sub{0}]
+\frac{2}{\D_-}\f_-^{-2}g\sub{0}^{ij}\pa_i\f_-\pa_j\f_-\right)\right)+c_-\chi$ \\&&\\ \hline
\end{tabular}
\end{center}
\caption{The renormalized effective actions corresponding to the $W_+$ and $W_-$ solutions in the
two-derivative approximation. $c_\pm$ are undetermined constants that depend on the dynamics, while
$\chi$ is the Euler number of the two-dimensional boundary. Note that the $d=2$ effective actions
are related to the Liouville action by the field redefinition $\vf=\log\f_-^{1/\D_-}$. Interestingly, the 
parameter $\x$ corresponds to the 2D cosmological constant. }
\label{ren-eff-actions}
\end{table}

The above analysis provides a complete rederivation of the constraints on the scalar mass in order for Dirichlet, Neumann
or Mixed boundary to be admissible \cite{Breitenlohner:1982bm,Balasubramanian:1998sn,Klebanov:1999tb}.
However, from a very different perspective. Here the constraints arise as essential conditions
for the {\em existence} of the corresponding effective action for the dual operator. As we have seen,
the existence of this effective action requires first of all the existence of a real function $W(\f)$,
which leads to the BF bound. As expected then, the $W_+$ solutions, corresponding to Dirichlet boundary conditions, lead
to an effective action for the VEV $\hp\sub{\D_+}=-\frac1l(\D_+-\D_-)\f_+$, for $\D_+>\D_-$, or $\hp\sub{\D_+}=-\frac1l\f_+$,
for $\D_+=\D_-=d/2$, of an operator of dimension $\D_+$. $W_-$ solutions on the other hand, which correspond to Neumann or
Mixed boundary conditions, lead to an effective action for the VEV $\f_-$ of an operator of dimension $\D_-$. It should
now be clear why $W_+$ solutions are isolated while $W_-$ solutions lie on a one-parameter family of $W_-$ solutions.
Namely, we have seen in Section \ref{bvp} that Dirichlet boundary conditions cannot be continuously deformed, but Neumann
and Mixed boundary conditions can. In particular, the parameter $\x$ that defines this one-parameter family of $W_-$
solutions is identified with the parameter of a marginal multi-trace deformation. Moreover, we have argued above that
these actions should only be valid away from the vanishing VEV point. Indeed, if the kinetic term in the actions in
Table \ref{ren-eff-actions} could be continued close to zero VEV, this would mean that the two-point function of the
scalar operator is dominated by the Goldstone pole in the UV (as well as in the IR), but this of course violates the
conformal invariance of the theory which should be restored in the UV. By looking at Table \ref{ren-eff-actions} we
see that the condition for the effective actions to break down for vanishing VEV is precisely the unitarity bound
$\D_\pm>(d-2)/2$. Since $\D_+\geq d/2$ by definition, this is only a constraint
on $\D_-$, which is equivalent to the condition that the mass lies in the range (\ref{mass_range}).
We conclude that Neumann and Mixed boundary conditions, which require the existence
of a $W_-$ solution, are only possible for masses in this range.\footnote{Note that this is in complete
agreement with the analysis of \cite{Amsel:2007im}, which shows that a $W_-(\f)$ solution is necessary
in order for stability with Mixed boundary conditions to be possible. But from our perspective in
terms of the dual field theory, the quantity $W(\f)$ (called $P(\f)$ in \cite{Amsel:2007im}) is physical
and not merely `an auxiliary construct' - it determines the effective action of the dual operator.}

\subsection{Minisuperspace approximation}

In order to see how a general multi-trace deformation can be accommodated in the Hamilton-Jacobi language, and
possibly to determine the parameters $c_\pm$ in the two-derivative effective actions in Table \ref{ren-eff-actions}
by computing the effective potential on a non-flat boundary, we
now proceed by considering the `minisuperspace' approximation for solutions of the form (\ref{generic-dw}). From
Table \ref{momenta} we see that with this ansatz the canonical momenta reduce to
$\p_A=-d(d-1)e^{dA}\dot{A}/\k^2$, $\p_\f=e^{dA}\dot{\f}$, while the Hamiltonian (\ref{Hamiltonian-min}) becomes
\be\label{red-H-min}
\ch=\frac12\left[\left(\p_\f^2-\frac{\k^2}{d(d-1)}\p_A^2\right)e^{-dA}-\left(-\frac{d(d-1)k}{\k^2l^2}e^{-2A}+2V(\f)\right)e^{dA}\right].
\ee
The Hamilton-Jacobi problem then reduces to a standard classical mechanics problem, where we look for a complete
integral, $\cs(A,\f)$, such that
\be\label{red-momenta}
\p_A=\frac{\pa \cs}{\pa A},\qquad \p_\f=\frac{\pa \cs}{\pa\f},
\ee
and $\ch=0$. For $k=0$, i.e. for a flat boundary, a solution to this Hamilton-Jacobi equation is
\be\label{complete-integral-1}
\cs(A,\f)=e^{dA}W(\f),
\ee
where $W(\f)$ satisfies equation (\ref{superpotential}). The two equations (\ref{red-momenta}) for the momenta then become respectively
\be\label{floweqs}
\dot{A}=-\frac{\k^2}{d-1}W(\f),\quad\dot{\f}=W'(\f).
\ee
In combination with equation (\ref{superpotential}) for the function $W(\f)$, we recognize
these equations as the flow or `BPS' equations for Poincar\'e domain walls (see e.g. \cite{ST,DeWolfe:1999cp}).
If the action (\ref{bulk_minimal_action}) is embedded into a particular gauged supergravity, then
generically there is a unique solution, $W_o(\f)$, of (\ref{superpotential}) that coincides with
the true superpotential of the theory. In that case the flow equations (\ref{floweqs}) do coincide with the true BPS
equations of the theory. However, any other solution $W(\f)$ gives a {\em non-supersymmetric} solution of the
supergravity equations \cite{Campos:2000yu, Papadimitriou:2006dr}. Following \cite{Freedman:2003ax}, we call $W(\f)$
the `fake superpotential', although `Hamilton's characteristic function' would be a more appropriate name in the
present context. Indeed, none of the above depends on supersymmetry in any way. The first order formalism, also
known as `fake supergravity' \cite{Freedman:2003ax}, for Poincar\'e domain walls is simply Hamilton-Jacobi theory
for the bulk equations of motion \cite{dBVV}. An analysis of curved domain walls, $k\neq 0$, in the context of
Hamilton-Jacobi theory has appeared recently in \cite{Skenderis:2006rr}. Since the Hamilton-Jacobi equation arising
from the Hamiltonian (\ref{red-H-min}) for $k\neq 0$ is non-separable, it is not easy to find a complete integral for
an arbitrary potential in this case. Here we will therefore focus on the $k=0$ case, but we will later show that for
a conformally coupled scalar such a complete integral can be found even for $k\neq 0$, which will allow us to fix the
constants $c_\pm$.

Recall that a complete integral of the Hamilton-Jacobi equation following from the Hamiltonian (\ref{red-H-min})
involves an arbitrary constant. In particular, the solution (\ref{complete-integral-1}) is a complete integral provided
$W(\f)$ is the general solution of (\ref{superpotential}), depending on an arbitrary parameter. This parameter, of course,
is the coupling, $\x$, of the marginal multi-trace deformation. Since any complete integral leads to the most general
solution of the form (\ref{generic-dw}), we can obtain the most general flat domain wall solution provided we can solve
equation (\ref{superpotential}) exactly for the one-parameter family of fake superpotentials. Indeed, we will show below
that this is possible, at least for certain potentials. However, the complete integral obtained form (\ref{complete-integral-1})
via the one-parameter family of fake superpotentials is not the most general complete integral. To see this, suppose $W_o(\f)$
is a solution of (\ref{superpotential}) such that (\ref{complete-integral-1}) is a solution (not a complete integral) of the
Hamilton-Jacobi equation. We can now look at the most general infinitesimal deformation, $\d\cs$, of this solution by
linearizing the Hamilton-Jacobi equation around the solution $\cs=e^{dA}W_o(\f)$. This gives
\be
W'_o(\f)\frac{\pa\d\cs}{\pa\f}-\frac{\k^2}{d-1}W_o(\f)\frac{\pa\d\cs}{\pa A}=0,
\ee
whose general solution is
\be\label{gen-def}
\d\cs=f\left(e^A e^{\left(\frac{\k^2}{d-1}\int^\f d\bar{\f}\frac{W_o}{W'_o}\right)}\right),
\ee
for an arbitrary function $f$. But note that asymptotically
$e^A e^{\left(\frac{\k^2}{d-1}\int^\f d\bar{\f}\frac{W_o}{W'_o}\right)}\sim \f_\pm^{1/\D_\pm}$, depending on
whether $W_o$ is a $W_+$ or $W_-$ solution. It follows that $\d\cs$ contributes to the renormalized action
and corresponds to an {\em arbitrary} multi-trace deformation. However, if this deformation were allowed
for both $W_+$ and $W_-$ solutions, it would contradict our previous conclusion that multi-trace deformations
are allowed only for Neumann or Mixed boundary conditions, and hence, only for $W_-$ solutions. Indeed, under
the deformation $\d\cs$,
\be
\d\dot{\f}\sim \frac{1}{\D_\pm}f'\left(e^A\f^{1/\D_\pm}\right)e^{-(d-1)A}\f^{(1-\D_\pm)/\D_\pm}.
\ee
In order for the asymptotic form of the scalar field, $\f\sim \f_\pm(x) e^{-\D_\pm}$, not to be changed by the
deformation, we must then require that $d-1+1-\D_\pm\geq\D_\pm$, or equivalently $\D_\pm\leq d/2$. But this
picks out only $\D_-$ since by definition $\D_+\geq d/2$. We therefore conclude that the deformation (\ref{gen-def})
is allowed only if $W_o$ is a $W_-$ solution.

The above discussion demonstrates that in the Hamilton-Jacobi formalism, any multi-trace deformation
corresponds to a choice of a complete integral of the Hamilton-Jacobi equation. In particular, although a complete
integral of the form (\ref{complete-integral-1}) accounts only for marginal multi-trace deformations, by allowing
for a more general complete integral as in (\ref{gen-def}) the Hamilton-Jacobi formalism can accommodate any
multi-trace deformation. This freedom in choosing a complete integral then gives rise to the arbitrary function
$f(\f_-)$ in the corresponding renormalized effective action.

\subsection{The `2/3' potential}

We now consider a special scalar potential for which equation (\ref{superpotential}) for the fake superpotential
can be solved exactly. As we have seen, this gives a complete integral of the Hamilton-Jacobi equation via (\ref{complete-integral-1}),
and hence the most general flat domain wall solution. The potential we will consider is
\be\label{2/3potential}
\framebox[2.7in]{
\begin{minipage}{3.9in}
\begin{equation*}
V(\f)=-\frac{d(d-1)}{2\k^2 l^2}\cosh\left(\frac23\sqrt{\frac{d\k^2}
{d-1}}\f\right),
\end{equation*}
\end{minipage}
}
\ee
which we propose to call the `$2/3$' potential. This potential was introduced in \cite{PS1},
although the special case $d=3$ has appeared elsewhere in the literature as well.
In particular, for $d=3$ this potential arises from a one-scalar consistent truncation of the $\cn=8$
gauged supergravity in $D=d+1=4$ dimensions \cite{Duff:1999gh}. It was also considered in \cite{MTZ},
where a four-dimensional asymptotically locally AdS topological black hole with scalar hair was found,
as well as in \cite{Papadimitriou:2006dr} and \cite{dHPP}, where respectively four-dimensional domain walls
and instantons were found and uplifted to M-theory.

The scalar mass for the potential (\ref{2/3potential}) is $m^2l^2=-2(d/3)^2$ and hence the two conformal
dimensions are $\D_-=d/3$, $\D_+=2d/3$. Requiring that the mass falls in the range (\ref{mass_range}), for
which Mixed boundary conditions can be considered, restricts the boundary dimension to lie in the range
$2\leq d\leq 6$. In \cite{PS1} it was shown that equation (\ref{superpotential}) with the potential (\ref{2/3potential})
can be solved exactly. The general solution is (see also \cite{Papadimitriou:2006dr})
\be\label{2/3-W}
W(\f;\n)=-\frac{d-1}{\k^2l}\frac{1}{(1-\r^2)^{\frac34}}\frac{1-\r^2+
\sqrt{1+2\n\r+\r^2}}{\sqrt{2(1+\n\r+\sqrt{1+2\n\r+\r^2})}},\quad \n\geq -1,
\ee
where $\r=\tanh\left(\frac23 \sqrt{\frac{d\k^2}{d-1}}\f\right)$ and $\n\geq -1$ is
an arbitrary parameter. In the $d=3$ case, the value $\n=-1$ corresponds to the true superpotential
of the truncated $\cn=8$ gauged supergravity in four dimensions \cite{Papadimitriou:2006dr}.
Expanding (\ref{2/3-W}) one obtains
\be\label{2/3-W-exp}
W(\f;\n)=-\frac{d-1}{\k^2l}\left(1+\frac16\psi^2+\frac{1}{27}\n\psi^3+\co(\psi^4)\right),
\ee
where $\psi=\sqrt{\frac{d\k^2}{d-1}}\f$. For any finite $\n\geq -1$ then
(\ref{2/3-W}) is a $W_-$ type solution and the parameter $\n$ is related to the
deformation parameter $\x$ in (\ref{def-exp}). In particular, choosing $\x$ such that $\x=0$
corresponds to the supersymmetric solution with $\n=-1$ gives
$\x=-\frac{d-1}{27\k^2l}\left(\frac{d\k^2}{d-1}\right)^{\frac32}(\n+1)$.

The flow equations (\ref{floweqs}) can now be used to construct the corresponding Poincar\'e
domain wall solution, which takes the form
\bea\label{2/3-dw}
ds^2&=&\left(\frac{3l}{d}\right)^2\frac{\left(1+\n\r+\sqrt{1+2\n\r+\r^2}
\right)}{2\r^2\sqrt{1-\r^2}(1+2\n\r+\r^2)}d\r^2+c^2\left(\frac{\sqrt{1-\r^2}}
{2\r^2}\left(1+\n\r+\sqrt{1+2\n\r+\r^2}\right)\right)^{3/d}dx^idx^i,\NO\\
\f&=&\frac32\sqrt{\frac{d-1}{d\k^2}}\tanh^{-1}\r,
\eea
where, for finite $\n$, the parameter $c$ is related to the VEV, $\f_-$, via
\be
c=\left(\frac23\sqrt{\frac{d\k^2}{d-1}}\f_-\right)^{3/d}.
\ee
For the special case $d=3$, where this domain wall is a solution of $\cn=8$ gauged
supergravity, the value $\n=-1$ gives a supersymmetric domain wall since for this value (\ref{2/3-W}) coincides
with the true superpotential. For $\n>-1$ (\ref{2/3-dw}) is a non-supersymmetric solution of the equations of motion.

We have seen that every solution of the bulk equations of motion is dual to an extremum, or `vacuum', of
the effective action of the dual boundary theory. In particular, Poincar\'e domain walls correspond to
homogeneous vacua, where the VEV $\f_-$ is a constant extremizing the effective potential $V_{\rm eff}(\f_-)$.
From our general prescription for computing the effective action in Section \ref{bvp}, we see that in
order to evaluate the effective potential we should first evaluate the renormalized momentum
as a functional of the VEV by solving the radial problem. We have already done this in this case
and we have found
$\hp\sub{\D_+}=d\x\f_-^{\D_+/\D_-}/\D_-=3\x\f_-^2$. However, this depends on the arbitrary parameter $\x$,
which should in principle be fixed by requiring regularity of the corresponding solution
(\ref{2/3-dw}). But since this solution is singular for any value of $\x$,\footnote{Note, however that
while for $\n=-1$ (\ref{2/3-dw}) has a null singularity, for $\n>-1$ the singularity is timelike
\cite{Papadimitriou:2006dr}. If this is taken as a criterion for `regularity', then the supersymmetric
solution is the only `regular' solution.} we {\em exceptionally} do not impose this condition and we will take $\x\leq0$
to be arbitrary. The effective potential then takes the form
\be
V_{\rm eff}(\f_-)=\x \f_-^{3}+f(\f_-),
\ee
where $\x$ is related to $\n$ by the relation we gave above. An extremum of this effective potential
then corresponds to imposing the relation
\be\label{extremum}
3\x \f_-^{2}+f'(\f_-)=0,
\ee
between the parameters $\n$ and $\f_-$. The domain wall (\ref{2/3-dw}) with this relation imposed is then
dual to the vacuum of the boundary theory corresponding to the VEV given by (\ref{extremum}). We should emphasize
here that given a solution of (\ref{extremum}), the corresponding  domain wall (\ref{2/3-dw}) takes the {\em same}
form for all choices of boundary condition $f(\phi_-)$, although the relation between $\nu$ and $\phi_-$ is different for different
boundary conditions. 

An interesting example is the case where the domain walls (\ref{2/3-dw}) are solutions of $\cn=8$
gauged supergravity in four dimensions. In that case, the $\n=-1$ ($\x=0$) domain wall is supersymmetric
and describes the Coulomb branch of the dual theory. The corresponding effective potential is therefore
flat, i.e. $f(\f_-)\equiv 0$, since the VEV is totally arbitrary. If the coupling $\x$ of the marginal
multi-trace deformation is then turned on, the effective potential, $V_{\rm eff}(\f_-)=\x \f_-^{3}$,
destabilizes the theory since $\x<0$.\footnote{If $\x$ were positive, the effective potential would force
the VEV to vanish. In that case the domain wall (\ref{2/3-dw}) reduces to exact AdS. A non-trivial solution
for $\n>-1$ then would require the addition of some other term proportional to $\x$ in the effective potential,
much like in the case $\x<0$.} To have a non-trivial solution when $\n>-1$ then, we need to introduce
another $\x$-dependent deformation, on top of the marginal one. In particular, we can choose $f(\f_-)=-\x h(\f_-)$
such that the effective potential is
\be
V_{\rm eff}(\f_-)=\x (\f_-^{3}-h(\f_-)),
\ee
for some function $h(\f_-)>0$. Even though for $\x=0$ the VEV is totally undetermined, when $\x$ is
turned on the VEV is fixed to some non-zero value determined by $V_{\rm eff}'(\f_-)=0$. This is precisely the
situation discussed in \cite{Papadimitriou:2006dr}, where the specific choice $h(\f_-)=J_*\f_-$, corresponding
to a {\em single}-trace deformation, was made. While at the supersymmetric point describing the Coulomb branch
the VEV is arbitrary, away from the supersymmetric point the VEV is a function of the arbitrary background
source, $J_*$, of the single-trace deformation.

Finally, let us consider the limit $\n\to \infty$, corresponding to the limit where the coupling, $\x$, of the
marginal deformation of the original CFT is sent to (negative) infinity. In this limit the fake superpotential
(\ref{2/3-W}) becomes
\be\label{2/3-W-lim}
W(\f;\infty)=-\frac{d-1}{\k^2l}\frac{1}{(1-\r^2)^{\frac34}}.
\ee
Expanding this we get
\be
W(\f;\infty)=-\frac{d-1}{\k^2l}\left(1+\frac13\psi^2+\co(\psi^4)\right),
\ee
and so this is a $W_+$ solution. The one-parameter family of fake superpotentials (\ref{2/3-W})
then provides an explicit example of the general picture we discussed in the previous section.
Namely, for all finite values of $\n$, (\ref{2/3-W}) is a $W_-$ solution, while the $W_+$ solution
arises as an endpoint of this one-parameter family at $\n\to\infty$. Letting $c^2\n^{\frac3d}=\bar{c}^2$,
the domain wall solution corresponding to $\n\to\infty$ is
\bea\label{2/3-dw-lim}
ds^2&=&\left(\frac{3l}{d}\right)^2\frac{d\r^2}{4\r^2\sqrt{1-\r^2}}
+\bar{c}^2\left(\frac{\sqrt{1-\r^2}}{2\r}\right)^{3/d}dx^idx^i,\NO\\
\f&=&\frac32\sqrt{\frac{d-1}{d\k^2}}\tanh^{-1}\r,
\eea
where now
\be
\bar{c}=\left(\frac43\sqrt{\frac{d\k^2}{d-1}}\f_+\right)^{3/2d}.
\ee
For an infinite value of the marginal deformation parameter then the dimension of the operator dual to the
scalar field changes from $\D_-$ to $\D_+$. The domain wall (\ref{2/3-dw-lim}) describes the arbitrary
VEV of this dimension $\D_+$ operator.

\section{Conformal coupling}
\label{conformal}
\setcounter{equation}{0}

As a second example of a system where the method outlined in Section \ref{method} can be applied, we
consider the minimally coupled scalar field in (\ref{bulk_minimal_action}) with the strange-looking potential
\be\label{conf-potential}
\framebox[6.5in]{
\begin{minipage}{6.9in}
\begin{equation*}
V(\f)=-\frac{d(d-1)}{2\k^2l^2}\left(\cosh\left(\sqrt{\frac{(d-1)\k^2}
{4d}}\f\right)\right)^{\frac{2(d+1)}{(d-1)}}+\frac{\l}{2}\left(
\frac{4d}{(d-1)\k^2}\right)^{\frac{(d+1)}{(d-1)}}\left(\sinh\left(
\sqrt{\frac{(d-1)\k^2}{4d}}\f\right)\right)^{\frac{2(d+1)}
{(d-1)}},
\end{equation*}
\end{minipage}
}
\ee
where $\l$ is an arbitrary dimensionless coupling constant, which we will assume it is positive. Although
this potential looks rather complicated and unintuitive, it is in fact a very special potential. First, note
that the scalar mass corresponding to the potential (\ref{conf-potential}) is the
conformal mass $m^2l^2=-(d/2)^2+1/4$, leading to the two conformal dimensions $\D_\pm=(d\pm 1)/2$.
Scalars with this mass in AdS are `massless' in the sense that their Lorentzian bulk-to-bulk
propagator has support only on the light cone $d(x,x')=0$, where $d(x,x')$ is the geodesic distance
between two points $x$ and $x'$ in AdS \cite{Breitenlohner:1982bm}. Moreover, the conformal mass falls
within the mass range (\ref{mass_range}) which allows for Mixed boundary conditions.

However, the conformal mass is not the only special property of the potential (\ref{conf-potential}).
Another special property of the potential (\ref{conf-potential}) is that for $d=3$ and $\l=\k^2/6l^2$ it
coincides with the potential (\ref{2/3potential}), which, as we pointed out, precisely for $d=3$ can be
embedded into $\cn=8$ gauged supergravity in four dimensions. The most significant property though of
(\ref{conf-potential}) is that the field redefinition
\be\label{field_redefs}
\sqrt{(d-1)/d}\k\tilde{\f}/2=\tanh\left(\sqrt{(d-1)/d}\k\f/2\right),\quad
\tilde{g}_{\m\n}=\left(\cosh\left(\sqrt{(d-1)/d}\k\f/2\right)
\right)^{\frac{4}{(d-1)}}g_{\m\n},
\ee
transforms the action (\ref{bulk_minimal_action}) with the potential (\ref{conf-potential})
into the form\footnote{Note that we have dropped the tildes from the action (\ref{bulk_conformal_action})
to simplify the formulas that follow. It should be clear from the context when $\f$ denotes
the minimally coupled scalar in (\ref{bulk_minimal_action}) or the conformally coupled scalar in
(\ref{bulk_conformal_action}).}
\bea\label{bulk_conformal_action}
S =\int_\cm d^{d+1}x \sqrt{g}\left(-\frac{1}{2\k^2}\left(R+\frac{d(d-1)}
{l^2}\right)+\frac12 g^{\m\n}\pa_\m\f\pa_\n\f
+\frac{d-1}{8d}R\f^2+\frac{\l}{2}\f^{\frac{2(d+1)}{(d-1)}}\right),
\eea
which is the action for a self-interacting scalar {\em conformally} coupled
to AdS$_{d+1}$ gravity. This transformation was given for the cases $d=2$ and
$d=3$ in \cite{Henneaux:2002wm} and \cite{MTZ} respectively. This last property allows
us to circumvent the problem of analyzing the action (\ref{bulk_minimal_action}) with
the complicated potential (\ref{conf-potential}), by studying instead the equivalent but
simpler action (\ref{bulk_conformal_action}). Note, however, that the transformation
(\ref{field_redefs}) implies that the conformally coupled scalar can be transformed into a
minimally coupled scalar provided it is bounded. Since the action (\ref{bulk_conformal_action})
does not necessarily imply that the scalar field is bounded, only certain bounded solutions
of (\ref{bulk_conformal_action}) correspond to solutions of (\ref{bulk_minimal_action}).

The equations of motion following from the action (\ref{bulk_conformal_action}) can be written in the form
\be\label{eoms}
R_{\m\n}+\frac{d}{l^2}g_{\m\n}=\k^2\ct_{\m\n},\quad
\square_g\f-\frac{d-1}{4d}R\f-\frac{d+1}{d-1}\l\f^{(d+3)/(d-1)}=0,
\ee
where the modified stress tensor $\ct_{\m\n}$ is given by
\bea\label{modified_stress_tensor}
\ct_{\m\n}=\frac{(d-1)^2}{4d}\frac{\f^{2d/(d-1)}}
{\left(1-\frac{(d-1)\k^2}{4d}\f^2\right)}
\left(\nabla_\m\nabla_\n-\frac{1}{d+1}g_{\m\n}\square_g\right)\f^{-2/(d-1)}.
\eea
These equations are in fact not independent. Since $\ct_{\m\n}$ is
manifestly traceless, the first equation in (\ref{eoms}) implies that the
Ricci scalar is constant
\be\label{Ricci}
R=-\frac{d(d+1)}{l^2}.
\ee
The contracted Bianchi identity then implies that $\ct_{\m\n}$ is
divergenceless. This fact imposes a differential constraint on the scalar
$\f$ which is  precisely the second equation in (\ref{eoms}), except that
the dimensionless coupling appears as an integration constant and so is not
determined by the divergencelessness of $\ct_{\m\n}$. Hence, the first
equation in (\ref{eoms}) implies the second up to the value of the
dimensionless coupling.

The very special form of these equations of motion makes it much easier to study
the action (\ref{bulk_conformal_action}) instead of the minimally coupled scalar described
by the action (\ref{bulk_minimal_action}) with the potential (\ref{2/3potential})- or any
other potential, in fact. However, the conformal coupling in the action (\ref{bulk_conformal_action})
requires that we revisit not only the variational problem, but also the derivation of the holographic
effective action. In Appendix \ref{appendix} we consider the variational problem for both the actions
(\ref{bulk_minimal_action}) and (\ref{bulk_conformal_action}) in detail, and in each case
we derive the correct form of the Gibbons-Hawking term, as well as the radial canonical
momenta, both of which are listed in Table \ref{momenta}. Using these results we can
now turn to the computation of the effective action for the operator dual to the scalar field
described by the action (\ref{bulk_conformal_action}).

\subsection{Two-derivative effective action for conformal boundary conditions}

To compute the renormalized effective action in the two-derivative approximation we proceed
as in the case of minimal coupling. Namely, one inserts the ansatz (\ref{action-ansatz})
for the regularized action into the momentum and Hamiltonian constraints (\ref{constraints}).
The momentum constraint is independent of the particular form of the canonical momenta and, as
in the case of minimal coupling, it is automatically satisfied since it simply reflects the
invariance of (\ref{action-ansatz}) with respect to $\S_r$ diffeomorphisms. Since the Hamiltonian
(\ref{Hamiltonian-conf}) is now different, however, the Hamiltonian constraint leads to the equations
\bea\label{Wconf}
&&W'^2-\frac{\k^2}{d(d-1)}\left(dW-\frac{d-1}{2}\f W'\right)^2
=-\frac{d(d-1)}{\k^2l^2}+\l\f^{\frac{2(d+1)}{d-1}}, \\\NO\\
\label{Zconf}
&&\left[W'+\frac{\k^2}{2d}\f\left(dW-\frac{d-1}{2}\f W'\right)\right]Z'
-\frac{\k^2}{d}\left(\frac{d-2}{d-1}\right)\left(dW-\frac{d-1}{2}\f W'\right)Z\NO\\
&&+\frac{1}{2\k^2}\left(1-\frac{(d-1)\k^2}{4d}\f^2\right)=0,\\\NO\\
\label{Mconf}
&&M=\frac{\frac{2\k^2}{d}\left(dW-\frac{d-1}{2}\f W'\right)Z'+\frac{d-1}{2d}\f}
{W'+\frac{\k^2}{2d}\f^2\left(dW-\frac{d-1}{2}\f W'\right)},
\eea
instead of equations (\ref{superpotential}), (\ref{Z}) and (\ref{M}). The action is
therefore determined once we solve the non-linear equation (\ref{Wconf}), which is
the analogue of (\ref{superpotential}) for minimal coupling.

The general solution of (\ref{Wconf}) in the vicinity of the exact solution corresponding to $\x=0$ takes the form
\be\label{Wconf-sol}
W(\f;\x)=-\frac{d-1}{\k^2l}+\left(\pm\frac{d-1}{2d}\sqrt{\l}+\x\left(1\mp\frac{2l\sqrt{\l}}{d-1}
\f^{\frac{2}{d-1}}\right)^{-d}\right)\f^{\frac{2d}{d-1}}+\co(\x^2),
\ee
where $\x$ is an arbitrary parameter analogous to the deformation parameter in (\ref{deformation}).
Note that both sign possibilities here lead to solutions analogous to the $W_-$ solutions of equation
(\ref{superpotential}) (i.e. $\f_-\neq 0$). Perhaps the analogue of a $W_+$ solution can be obtained in
the limit $\x\to\infty$, which could be evaluated if the solution (\ref{Wconf-sol}) were known exactly as
a function of $\x$, but we will not investigate this further. We will fix the sign in (\ref{Wconf-sol})
below. Using (\ref{Wconf-sol}) we find that the leading asymptotic behavior of the functions $Z(\f)$ and
$M(\f)$ is exactly as given in Table \ref{ZM-asymptotics} for a $W_-$ solution with $\D_-=(d-1)/2$.
Evaluating then the renormalized action we
obtain
\bea\label{conf-ren-eff-action}
\G[\f_-]=\left\{\begin{matrix} \int_{\pa\cm}d^dx\sqrt{g\sub{0}}\left(V_{\rm eff}(\f_-)+c_-\left(\f_-^{\frac{2(d-2)}{(d-1)}}R[g\sub{0}]
+\frac{4(d-2)}{(d-1)}\f_-^{-\frac{2}{(d-1)}}g\sub{0}^{ij}\pa_i\f_-\pa_j\f_-\right)\right), & d>2,\\&\\
\int_{\pa\cm}d^2x\sqrt{g\sub{0}}\left(V_{\rm eff}(\f_-)+\frac{l}{\k^2}\left(\log\f_-R[g\sub{0}]
+4\f_-^{-2}g\sub{0}^{ij}\pa_i\f_-\pa_j\f_-\right)\right)+c_-\chi, & d=2,\\
\end{matrix}\right.
\eea
which, as expected, are identical with the renormalized actions in Table \ref{ren-eff-actions}, except that the effective potential is now given by
\be\label{conf-flat-eff-potential}
V_{\rm eff}(\f_-)=\left(\pm\frac{d-1}{2d}\sqrt{\l}+\x\right)\f_-^{\frac{2d}{d-1}}+f(\f_-).
\ee

\subsection{Minisuperspace approximation}

As for the minimally coupled scalar, the next step is to look at the `minisuperspace' of solutions of the
form (\ref{generic-dw}). The canonical momenta dual to the warp factor, $A(r)$, and the scalar field,
which can be deduced from the momenta given in Table \ref{momenta}, are respectively
\bea\label{conf-mom}
&&\p_A=e^{dA}\left(-\frac{d(d-1)}{\k^2}\left(1-\frac{(d-1)\k^2}{4d}\f^2\right)\dot{A}+\frac{d-1}{2}\f\dot{\f}
\right),\NO\\
&&\p_\f=e^{dA}\left(\dot{\f}+\frac{d-1}{2}\dot{A}\f\right).
\eea
Moreover, the Hamiltonian (\ref{Hamiltonian-conf}) becomes
\bea\label{red-H-conf}
\ch&=&\frac12\left\{e^{-dA}\left(\p_\f^2-\frac{\k^2}{d(d-1)}\left(\p_A-\frac{d-1}{2}\f\p_\f\right)^2\right)
\right.\NO\\
&&\left.+e^{dA}\left(\frac{d(d-1)}{\k^2l^2}-\l\f^{\frac{2(d+1)}{(d-1)}}+\frac{d(d-1)k}{\k^2l^2}e^{-2A}
\left(1-\frac{(d-1)\k^2}{4d}\f^2\right)\right)\right\}.
\eea
Writing again
\be\label{red-momenta-conf}
\p_A=\frac{\pa \cs}{\pa A},\qquad \p_\f=\frac{\pa \cs}{\pa\f},
\ee
and inserting these into the equation $\ch=0$ for the Hamiltonian (\ref{red-H-conf}) we
obtain the Hamilton-Jacobi equation for the conformally coupled scalar.

We could now look for a solution of the form (\ref{complete-integral-1}), in which case the Hamilton-Jacobi
equation requires that the fake superpotential satisfies equation (\ref{Wconf}). Indeed, the exact
solution obtained from (\ref{Wconf-sol}) by setting $\x=0$ does give a solution to the Hamilton-Jacobi equation.
However, since we do not know the full one-parameter family of fake superpotentials that solve (\ref{Wconf}),
the corresponding solution of the Hamilton-Jacobi equation is not a complete integral, which is necessary
in order to obtain the most general domain wall solutions of the equations of motion. Nevertheless, in this
case we can find a complete integral of the Hamilton-Jacobi equation that is {\em not} of the form (\ref{complete-integral-1})
and it is valid even for curved boundary, $k=\pm1$, as well as for flat boundary. It is easy to verify that writing
\be\label{momenta-sol}
\framebox[5.2in]{
\begin{minipage}{5.5in}
\begin{eqnarray*}
\p_A&=&-\frac{d(d-1)}{\k^2l}e^{dA}\sqrt{1+ke^{-2A}+\m e^{-(d+1)A}}+\frac{d-1}{2}\f\p_\f,\NO\\\NO\\
\p_\f&=&\pm e^{dA}\sqrt{\frac{d(d-1)}{\k^2l^2}\m e^{-(d+1)A}
+k\left(\frac{d-1}{2l}\right)^2\f^2 e^{-2A}+\l\f^{\frac{2(d+1)}{(d-1)}}},
\end{eqnarray*}
\end{minipage}
}
\ee
where $\m$ is an arbitrary parameter, in the Hamiltonian (\ref{red-H-conf}) automatically gives $\ch=0$.
Of course, this does not mean that we have found a solution to the Hamilton-Jacobi equation unless
\be
\frac{\pa\p_A}{\pa\f}=\frac{\pa\p_\f}{\pa A}.
\ee
Remarkably, this is indeed the case and hence there exists a complete integral $\cs(A,\f)$ such that
the momenta (\ref{momenta-sol}) are obtained from it via (\ref{red-momenta-conf}). The fact that
the momenta (\ref{momenta-sol}) are integrable, i.e. that they can be derived from a complete integral
$\cs(A,\f)$ via (\ref{red-momenta-conf}), is one of our main results. As we will now show, this
complete integral of the Hamilton-Jacobi equation will allow us not only to completely determine
the two-derivative effective action on any boundary of locally constant scalar curvature, but also
to obtain all possible solutions of the form (\ref{generic-dw}).

\begin{flushleft}
{\bf Effective action}
\end{flushleft}

Expanding the momenta (\ref{momenta-sol}) in eigenfunctions of the dilatation operator and
keeping the term of weight zero gives immediately the renormalized momenta
\bea\label{ren-mom-exact}
&&\hp\sub{d}^i_j=\frac12(1+(-1)^d)\frac{(-1)^{\frac d2}k^{\frac d2}\G(d+1)}{2^{d+1}l\k^2
\left(\G\left(\frac d2+1\right)\right)^2}\d^i_j+\frac{d-1}{4d}\f_-\hp\sub{\D_+}\d^i_j,\NO\\\NO\\
&&\hp\sub{\D_+}=\pm\sqrt{\frac{d(d-1)}{\k^2l^2}\m+k\left(\frac{d-1}{2l}\right)^2\f_-^2
+\l\f_-^{\frac{2(d+1)}{(d-1)}}},
\eea
where we have traded again the momentum $\p_A$ for the physical momentum conjugate to the induced metric $\g_{ij}$.
Note that the first term in $\hp\sub{d}^i_j$, which only appears for even boundary dimension, is nothing but the
conformal anomaly \cite{HS}, as can be seen from the relation between the VEV of the stress tensor and the renormalized
momentum $\hp\sub{d}^i_j$ given in Table \ref{dictionary}.

The first thing that these renormalized momenta can tell us is the value of the undetermined parameter $c_-$
in the effective action (\ref{conf-ren-eff-action}). Note that the parameter $\m$ is determined
as a function of the VEV $\f_-$ by the requirement of regularity for the corresponding domain wall,
which we will discuss below. As we will show, a possible value is $\m=0$ - in fact the only possible value
for the physically relevant case where (\ref{bulk_conformal_action}) is embedded in $\cn=8$
gauged supergravity in four dimensions. Choosing $\m=0$ then and expanding $\hp\sub{\D_+}$ for small curvature (large $l$), we obtain
\be
\hp\sub{\D_+}=\pm\sqrt{\l}\f_-^{\frac{d+1}{d-1}}\left(1+\frac{k}{2\l}\left(\frac{d-1}{2l}\right)^2
\f_-^{-\frac{4}{d-1}}+\cdots\right),
\ee
where the dots stand for higher derivative terms. Comparing this with the derivative of the
effective action (\ref{conf-ren-eff-action}) with respect to the VEV, $\f_-$, and using $R[g\sub{0}]=d(d-1)/l^2$, determines
\be\label{c}
\framebox[4.in]{
\begin{minipage}{5.in}
\begin{equation*}
c_-=\pm\frac{(d-1)^2}{16d(d-2)\sqrt{\l}}, \quad d>2, \qquad \sqrt{\l}=\k^2/16l, \quad d=2.
\end{equation*}
\end{minipage}
}
\ee
Note in particular that for $d=2$ the coupling $\l$ is itself fixed. In order to have a positive kinetic
term in the effective action (\ref{conf-ren-eff-action}), we should choose the positive sign in the renormalized
momentum $\hp\sub{\D_+}$ and hence in the momentum (\ref{momenta-sol}).

However, the renormalized momentum $\hp\sub{\D_+}$ in (\ref{ren-mom-exact}) allows us to determine the exact
effective potential on any boundary of constant scalar curvature. Namely, integrating $\hp\sub{\D_+}$ with
respect to $\f_-$ (for $\m=0$ again) we obtain the exact effective potential
\be\label{exact_effective_potential_k}
\framebox[5.5in]{\rule[-.3in]{0in}{0in}
\begin{minipage}{6.in}
\begin{eqnarray*}
V_k(\f_-)&=&\frac{(d-1)^3k}{8d(d-2)l^2\sqrt{\l}}\f_-^{\frac{2(d-2)}{(d-1)}}F\left(1-\frac{d}{2},\frac12;2
-\frac{d}{2};-\frac{k}{\l}\left(\frac{d-1}{2l}\right)^2\f_-^{-\frac{4}{(d-1)}}\right)\NO\\
&&+\frac{(d-1)}{2d}\f_-\sqrt{\left(\frac{d-1}{2l}\right)^2 k\f_-^2+
\l\f_-^{\frac{2(d+1)}{(d-1)}}}+V_o+f(\f_-),
\end{eqnarray*}
\end{minipage}
}
\ee
where the overall constant
\be
V_o=-\frac{\G\left(2-\frac d2\right)\G\left(\frac{d-1}{2}\right)}{\G\left(\frac12\right)}
\frac{(d-1)^{d+1}k^{\frac d2}}{2^{d+1}d(d-2)l^d\l^{\frac{d-1}{2}}},
\ee
is determined by the requirement that $V_k(0)=0$. For $k=0$ the effective potential  (\ref{exact_effective_potential_k}) reduces to
\be\label{exact_effective_potential}
V_{k=0}(\f_-)=\frac{(d-1)}{2d}\sqrt{\l}\f_-^{\frac{2d}{(d-1)}}+f(\f_-).
\ee
For $k=1$ and for $d=2,3,4$ the potential (\ref{exact_effective_potential_k})
is explicitly shown in Table \ref{effective-potentials}.

\begin{table}
\begin{center}
\begin{tabular}{|c|c|}
\hline &\\
$d$ & $V_{k=1}(\f_-)$ \\&\\\hline&\\
2 & $\frac{1}{8l}\left(\f_-^2\sqrt{1+4l^2\l\f_-^4} +\frac{1}{2l\sqrt{\l}}
\log\left(2l\sqrt{\l}\f_-^2+\sqrt{1+4l^2\l\f_-^4}\right)\right)+f(\f_-)$ \\&\\
3 & $\frac{1}{3l^3\l}\left(\left(1+l^2\l\f_-^2\right)^{3/2}-1\right)+f(\f_-)$\\&\\
4 & $\frac{3}{8\l}\left(\frac{3}{2l}\right)^3\left(\f_-^{2/3}\left(\frac12+\frac{4l^2}{9}\l\f_-^{4/3}\right)
\sqrt{1+\frac{4l^2}{9}\l\f_-^{4/3}}-\frac{1}{4l\sqrt{\l}}\log\left(\frac{2l}{3}\f_-^{2/3}+
\sqrt{1+\frac{4l^2}{9}\l\f_-^{4/3}}\right)\right)+f(\f_-)$\\&\\\hline
\end{tabular}
\end{center}
\caption{The exact effective potential (\ref{exact_effective_potential_k}) for $k=1$ and
$d=2,3,4$. Note that for $d=2$ the coupling, $\l$, is given by (\ref{c}).}
\label{effective-potentials}
\end{table}

\begin{flushleft}
{\bf Domain walls and (absence of) gravitational instantons}
\end{flushleft}

The solution (\ref{momenta-sol}) also enables us to find all possible solutions of the form
(\ref{generic-dw}). Using the expressions (\ref{conf-mom}) for the canonical momenta in terms
of the radial derivatives of the warp factor and of the scalar field we find
\be\label{warp-sol}
dr=\frac{ldA}{\sqrt{1+ke^{-2A}+\m e^{-(d+1)A}}},
\ee
where $\m$ is an arbitrary integration constant. Defining $u\equiv e^{-A}$ and $\vf\equiv u^{-(d-1)/2}\f$
we can then write down the most general domain wall (\ref{generic-dw}) in the form
\be\label{conf-dw-sol}
\framebox[4.7in]{\rule[-.3in]{0in}{0in}
\begin{minipage}{6.in}
\begin{eqnarray*}
&&ds^2=\frac{l^2du^2}{u^2\left(1+ku^2+\m u^{d+1}\right)}+
\frac{1}{u^2}ds_d^2.\\\\
&&\int_{\vf_o}^\vf\frac{d\bar{\vf}}{\sqrt{\frac{d(d-1)}{\k^2l^2}\m+k\frac{(d-1)^2}{4l^2}
\bar{\vf}^2+\l\bar{\vf}^{\frac{2(d+1)}{(d-1)}}}}=\mp l\int_0^u\frac{d\bar{u}}
{\sqrt{1+k\bar{u}^2+\m \bar{u}^{d+1}}}.
\end{eqnarray*}
\end{minipage}
}
\ee
The slice metric $ds_d^2$ is either one of the three constant curvature metrics given in Table
\ref{slice-metrics} or the metric on a quotient of these by a discrete subgroup of their isometry
group. Moreover, since we picked the plus sign in the renormalized momentum $\hp\sub{\D_+}$ in
(\ref{ren-mom}) so that the kinetic term in the effective action has a positive sign, we must
pick the {\em minus} sign in (\ref{conf-dw-sol}).
\begin{table}
\begin{center}
\begin{tabular}{|c|l|l|}
\hline &&\\
$k$ & $ds_d^2$ & \\&&\\\hline&&\\
0 & $ds^2(\mathbb{R}^d)=dx^idx^i$ & $z=u$, $z^i=x^i$ \\&&\\
1 & $ds^2(S^d)=d\th^2+\sin^2\th ds^2(S^{d-1})$& $z=\frac{u}{\sqrt{1+u^2}+\cos\th}$, $z^i=\frac{\sin\th n^i}{\sqrt{1+u^2}+\cos\th}$, $n^in^i=1$\\&&\\
-1 & $ds^2(\mathbb{H}_d)=d\th^2+\sinh^2\th ds^2(S^{d-1})$ & $z=\frac{u}{\sqrt{1-u^2}+\cosh\th}$,
$z^i=\frac{\sinh\th n^i}{\sqrt{1-u^2}+\cosh\th}$, $n^in^i=1$\\&&\\\hline
\end{tabular}
\end{center}
\caption{The three slice metrics $ds_d^2$ corresponding respectively to $k=0,\pm1$, and
the coordinate transformations that bring the metric (\ref{conf-dw-sol}) with $\m=0$ into
the upper half plane metric (\ref{uhp-coords}).}
\label{slice-metrics}
\end{table}

Note that if either $k$ or $\m$ are negative, then the range of the radial coordinate is bounded,
$0\leq u\leq u_*$, for some upper bound $u_*$.\footnote{In fact, for $k=-1$ and $\m>0$ these solutions
are very similar to the Janus solution \cite{Bak:2003jk}.} For $k$ and $\m$ non-negative, however, $u$
is unbounded from above: $0\leq u<\infty$. Since $\f=u^{(d-1)/2}\vf$, in this case regularity of the solution
(\ref{conf-dw-sol}) requires that $\vf\to 0$ as $u\to\infty$.\footnote{Note, however, that regularity
of the solutions for the conformally coupled scalar (\ref{bulk_conformal_action}) does not
guarantee the regularity of the corresponding minimally coupled scalar since the transformation
rules (\ref{field_redefs}) may break down. This should be checked case by case.} This gives the condition
\be
\int_{\frac{d-1}{2l\sqrt{\l}}\vf_o^{-2/(d-1)}}^\infty\frac{dv}{\sqrt{1+kv^2
+\frac{d(d-1)}{\k^2l^2\l}\left(\frac{2l\sqrt{\l}}{d-1}\right)^{d+1}\m v^{d+1}}}=
\int_0^\infty\frac{dv}{\sqrt{1+kv^2+\m v^{d+1}}}.
\ee
For $\m=0$ this is trivially satisfied since both integrals diverge. It follows that $\m=0$
leads to regular solutions, which we will discuss shortly. For $\m>0$, however, this
constraint can only be satisfied if
\be
\l< \left(\frac{(d-1)^d\k^2}{2^{d+1}d l^{d-1}}\right)^{2/(d-1)}.
\ee
Provided this holds, the above constraint gives a relation between $\m$ and the VEV $\f_-$. Note, however,
that for $d=2$ and $d=3$ the condition on the coupling is respectively $\l<(\k^2/16l)^2$ and $\l<\k^2/6l^2$.
But recall that we have determined in (\ref{c}) that for $d=2$ we must necessarily have $\l=(\k^2/16l)^2$,
while for $d=3$,  $\l=\k^2/6l^2$ is precisely the value of the coupling such that the
action (\ref{bulk_conformal_action}) can be embedded in M-theory. It follows that no
regular solutions of the form (\ref{conf-dw-sol}) with both $k$ and $\m$ non-negative
exist in these two cases. This is particularly significant for the three-dimensional case
that can be embedded in M-theory. Note that (\ref{conf-dw-sol}) with $k=1$ and $\m>0$, were it a
regular solution, it would be a gravitational instanton similar to the numerical solutions of \cite{Hertog:2004rz}.
As in that case, by analytic continuations (\ref{conf-dw-sol}) would then give
static spherically symmetric gravitational solitons, as well as Big Bang/Crunch geometries,
which would have a dual description in the dual $\cn=8$ strongly coupled SCFT in three dimensions.
Our argument shows the absence of such solitons in this theory. Of course, this refers to
the particular truncation of $\cn=8$ gauged supergravity that gives the action
(\ref{bulk_conformal_action}), and which is {\em different} from the one used in \cite{Hertog:2004rz}.
We will see in the next section, however, that (\ref{bulk_conformal_action}) does admit
regular instanton solutions, but of a different type.

An interesting property of the general domain wall solution (\ref{conf-dw-sol}) is that for
$\m=0$ the metric is the metric of Euclidean AdS$_{d+1}$ (i.e. of the hyperbolic space
$\mathbb{H}_{d+1}$), for all possible values of $k=0,\pm1$.\footnote{Of course, this is true
provided there are no global identifications in the slice metric $ds_d^2$.} Indeed, the coordinate
transformations given in Table \ref{slice-metrics} for each of the cases $k=0,\pm1$, transform
the metric in (\ref{conf-dw-sol}) with $\m=0$ to the upper half plane metric of $\mathbb{H}_{d+1}$
\be\label{uhp-coords}
ds^2=\frac{l^2}{z^2}(dz^2+dz^idz^i),\quad i=1,\ldots,d.
\ee
Moreover, for $\m=0$ the scalar field in (\ref{conf-dw-sol}) takes the form
\be\label{mu=0}
\f=\vf_o u^{\frac{d-1}{2}}\left(\sqrt{1+ku^2}\pm u\sqrt{k+\left(2l\sqrt{\l}\vf_o^\frac{2}{d-1}/(d-1)\right)^2}\right)^{-(d-1)/2},
\ee
where $\vf_o$ is an arbitrary constant corresponding to the VEV $\f_-$. For $k\geq 0$, $\vf_o\geq0$ but
for $k=-1$, $\vf_o>((d-1)l/\sqrt{2\l})^{(d-1)/2}$. Note also that with the sign choice we made above,
the plus sign should be chosen in this expression. We will revisit this special case of the domain wall solutions (\ref{conf-dw-sol}) shortly.

Another special case of the general solution (\ref{conf-dw-sol}) deserves a comment. Namely,
for $d=3$ and $\l=\k^2/6l^2$, in which case the two potentials (\ref{2/3potential}) and
(\ref{conf-potential}) agree, the Poincar\'e domain wall (\ref{2/3-dw}) that we found in the previous
section matches precisely with the domain wall (\ref{conf-dw-sol}) for $k=0$.
In order to compare  the two solutions we use the field redefinitions (\ref{field_redefs}) to
rewrite the domain wall (\ref{2/3-dw}) in the frame where the scalar field is conformally coupled. The transformed solution takes the form
\bea\label{dw2-conf}
ds^2&=&\frac{1}{4\r^2}\left(1+\n\r+\sqrt{1+2\n\r+\r^2}\right)\left(1+\sqrt{1-\r^2}\right)
\left(\frac{l^2d\r^2}{(1-\r^2)(1+2\n\r+\r^2)}+\r_o^2dx^idx^i\right),\NO\\
\f&=&\frac{\sqrt{6}}{\k}\frac{\r}{1+\sqrt{1-\r^2}},
\eea
where $\r_o=\frac{2\k}{\sqrt{6}}\f_-$. It is now straightforward to check that identifying
\be
u^{-2}=\frac{\r_o^2}{4\r^2}\left(1+\n\r+\sqrt{1+2\n\r+\r^2}\right)\left(1+\sqrt{1-\r^2}\right),
\ee
the solution (\ref{dw2-conf}) reproduces the domain wall (\ref{conf-dw-sol}),
provided we also identify $\m=(\n^2-1)\r_o^4/16$. According to the above discussion then,
only the supersymmetric domain wall, corresponding to $\n=-1$, is regular. However, even in this
case, the corresponding solution for the physically relevant minimally coupled scalar is not
regular. As was discussed in \cite{Papadimitriou:2006dr}, the difference between the
supersymmetric and non-supersymmetric domain walls is that the former has a null singularity,
while the later has a timelike singularity. In both cases this singularity can be seen
in the corresponding 11-dimensional solution as arising from a collapsing $S^2$ inside the
asymptotically $S^7$ transverse space.

\subsection{Instantons}

Finally, we consider a very special class of solutions of the equations of motion (\ref{eoms}), which
for the case $d=3$ were found in \cite{dHPP}. These are solutions that satisfy
\be\label{instanton-eqs}
\ct_{\m\n}=0,\qquad R_{\m\n}+\frac{d}{l^2}g_{\m\n}=0.
\ee
In the case of a vanishing cosmological constant such solutions have been
discussed in \cite{Ayon-Beato:2005tu}. The vanishing of the modified
stress tensor gives a {\em linear} equation for the scalar field, namely
\be\label{linear_eq}
\left(\nabla_\m\nabla_\n-\frac{1}{d+1} g_{\m\n}\square_g\right)\f^{-2/(d-1)}=0,
\ee
which admits non-trivial solutions provided the metric is exact AdS$_{d+1}$.
The general solution of this equation, subject to the constraint that it also
satisfies the second equation in (\ref{eoms}), can be written in the upper half plane coordinates
(\ref{uhp-coords}) as
\be\label{instanton_uhp}
\framebox[3.7in]{
\begin{minipage}{5.in}
\begin{equation*}
\f^{2/(d-1)}=\frac{(d-1)}{l\sqrt{|\l|}}\left(\frac{bz}{-{\rm sgn}(\l)b^2
+(z+a)^2+(\vec{z}-\vec{z}_0)^2}\right),
\end{equation*}
\end{minipage}
}
\ee
where $a,b,z_0^i$, $i=1,\ldots,d$, are arbitrary constants.
A special case of this solution (for $\l<0$) was found in \cite{dH&P} as a solution of the
scalar equation (\ref{eoms}) in four dimensions and ignoring the back-reaction on the geometry.
It was later pointed out in \cite{dHPP} that for any value of the coupling, $\l$, there is in fact no
back-reaction and, together with the AdS$_4$ metric, this is an exact solution of the full
gravity-scalar system. The solution (\ref{instanton_uhp}) is the generalization of the exact solution of \cite{dHPP} to any dimension.

In order to understand the significance of the parameters in this solution we consider its
asymptotic expansion
\be\label{asymptotics}
\f\sim e^{-\frac{(d-1)r}{2l}}\f_-(\vec{z})-l\a e^{-\frac{(d+1)r}{2l}}
\f_-^{\frac{(d+1)}{(d-1)}}(\vec{z}),
\ee
where $\a\equiv \sqrt{|\l|}a/b$ and the inhomogeneous VEV, $\f_-(\vec{z})$, is given by
\be\label{VEV}
\f_-^{2/(d-1)}=\frac{(d-1)}{l\sqrt{|\l|}}\left(\frac{b}{-{\rm sgn}(\l)b^2
+a^2+(\vec{z}-\vec{z}_0)^2}\right).
\ee
The asymptotic form (\ref{asymptotics}) tells us that the two modes, $\f_\pm$, of this soliton
are related by $\f_+=-l\a\f_-^{\frac{(d+1)}{(d-1)}}$, or equivalently
\be
\hp\sub{\D_+}=-\frac1l(\D_+-\D_-)\f_+=\a\f_-^{\frac{d+1}{2}}.
\ee
From Table \ref{bcs} we then unambiguously conclude that the solution (\ref{instanton_uhp})
satisfies Mixed boundary conditions: $J_{f_-}=-\hp\sub{\D_+}-f'(\f_-)=0$, with
\be\label{conformal_deformation}
f(\f_-)=-\a\frac{(d-1)}{2d}\f_-^{\frac{2d}{(d-1)}},
\ee
which corresponds to a {\em marginal} multi-trace deformation with deformation parameter $\a$.
It follows that the parameter $\a=\sqrt{|\l|}a/b$ is not a modulus of the solution (i.e. of the VEV),
but rather a modulus of the {\em theory} itself (of course this refers to the large $N$ limit only).
Different values of $\a$ correspond to different points along the line of marginal deformations (\ref{conformal_deformation}).
Equivalently, two solutions of the form (\ref{instanton_uhp}) with different values of $\a$ satisfy {\em different} boundary
conditions. Note that regularity of the general solution
(\ref{instanton_uhp}) requires that $a>b\geq 0$ and hence $\a>\sqrt{\l}>0$. Using the exact
effective potential (\ref{exact_effective_potential}) with the boundary condition
(\ref{conformal_deformation}) we see that this is precisely the condition for the effective potential to
become unbounded from below.\footnote{Putting the theory on $S^d$ does not change this conclusion, as can
be deduced from the effective potential (\ref{exact_effective_potential_k}).} This suggests that the
Euclidean solutions (\ref{instanton_uhp}) are instantons which mediate the decay of the conformal vacuum
at $\f_-=0$ due to the instability introduced by the marginal deformation (\ref{conformal_deformation}) \cite{dHPP}.

A curious feature of the VEV (\ref{VEV}) is that it is an extremum of a simple two-derivative
boundary action \cite{dHPP}.\footnote{See \cite{Zamolodchikov:2006xs} for a recent discussion
of the $d=2$ case.} This `phenomenological' effective action in flat space takes the form
\be\label{pheno}
S=\cc\int d^dz \left(\f_-^{-\frac{2}{d-1}}\pa^i\f_-\pa_i\f_-+(\l-\a^2)\f_-^{\frac{2d}{d-1}}\right),
\ee
where $\cc$ is an arbitrary constant. Note, however, that this effective action does {\em not} -
and indeed it does not have to - agree with the holographic two-derivative effective action
(\ref{conf-ren-eff-action}) we have derived above. If one chooses $\cc$ such that it matches
the correct coefficient of the kinetic term given in (\ref{conf-ren-eff-action}), then the
coefficient of the potential term in (\ref{pheno}) should be changed as
$\l-\a^2\to2\sqrt{\l}(\sqrt{\l}-\a)$, if  (\ref{pheno}) were to agree with (\ref{conf-ren-eff-action}).
Note that in the vicinity of the critical point at $\a=\sqrt{\l}$ these coefficients {\em do}
actually agree. However, we see this as merely a curious coincidence, since the correct two-derivative
effective action as we showed is (\ref{conf-ren-eff-action}), and indeed the instanton VEV
(\ref{VEV}), which is an exact extremum of the full {\em all}-derivative effective action,
need not be an extremum of the two-derivative effective action.

\begin{flushleft}
 {\bf Moduli space}
\end{flushleft}

The moduli space of the instantons (\ref{instanton_uhp}) is the space parameterized by all
arbitrary parameters of the solution, subject to the condition that the boundary conditions
remain fixed, i.e. provided $\a$ remains fixed. The moduli space becomes manifest if we
rewrite the instanton solution (\ref{instanton_uhp}) in terms of the coordinates
$(Y_{-1},Y_0,Y_i)$, $i=1,\ldots,d$, of the covering space, $\mathbb{R}^{1,d+1}$, of $\mathbb{H}_{d+1}$.
Namely, we introduce coordinates on  $\mathbb{R}^{1,d+1}$ as well as a set of auxiliary constants
parameterizing an $\widetilde{\mathbb{H}}_{d+1}$ as
\be
\begin{array}{ll}
Y_{-1}=\frac{l}{2z}(1+z^2+\vec{z}^2), & \tilde{Y}_{-1}=\frac{\tilde{l}}
{2\tilde{z}}(1+\tilde{z}^2+\vec{\tilde{z}}^2),\\\\
Y_0=\frac{l}{2z}(1-z^2-\vec{z}^2),& \tilde{Y}_0=\frac{\tilde{l}}{2\tilde{z}}
(1-\tilde{z}^2-\vec{\tilde{z}}^2), \\ \\
Y_i=\frac{l}{z}z^i, & \tilde{Y}_i=\frac{\tilde{l}}{\tilde{z}}\tilde{z}^i.
\end{array}
\ee
The solution (\ref{instanton_uhp}) then can be written as
\be\label{instanton_covering}
\framebox[2.in]{
\begin{minipage}{5.in}
\begin{equation*}
\f^{2/(d-1)}=\frac{1}{-\tilde{Y}\cdot Y+\frac{2}{d-1}l\a},
\end{equation*}
\end{minipage}
}
\ee
where $\tilde{Y}\cdot Y\equiv -\tilde{Y}_{-1}Y_{-1}+\tilde{Y}_0Y_0+\sum_{i=1}^d
\tilde{Y}_iY_i$, and we have identified
\be
\tilde{z}=\frac{(d-1)}{2}\frac{b\tilde{l}}{\sqrt{|\l|}},\quad
\tilde{z}^i=z_o^i,\quad\tilde{l}=\frac{2}{d-1}\sqrt{\a^2-\l}.
\ee
The moduli space is therefore a hyperbolic space, $\tilde{\mathbb{H}}_{d+1}$,
of radius $\tilde{l}$, which is well defined precisely for $\a>\sqrt{\l}$.
Recall that this is exactly the condition for the instantons to exist, as well as,
for the effective potential to be unbounded from below. Finally, note that the form
(\ref{instanton_covering}) allows one to easily write the solution in any other
coordinate system parameterizing $\mathbb{H}_{d+1}$.

\begin{flushleft}
{\bf Special limits}
\end{flushleft}

Since the metric (\ref{conf-dw-sol}) becomes exact AdS$_{d+1}$ for $\m=0$ and (\ref{instanton_uhp}) is
the most general solution corresponding to an exact AdS$_{d+1}$ metric, it follows that we must
be able to obtain the $\m=0$ domain wall solutions discussed above as a limit of the solution
(\ref{instanton_uhp}). This can indeed be easily seen, provided we realize that in taking any limits of the
general solution (\ref{instanton_uhp}), we do not necessarily have to satisfy the
the condition $a>b\geq 0$ that was essential for the regularity of the general solution. This
is especially so since the special limits can satisfy more general boundary conditions
than the general solution (\ref{instanton_uhp}).

\begin{figure}
\begin{center}
\scalebox{1.0}{\rotatebox{-0}{\includegraphics{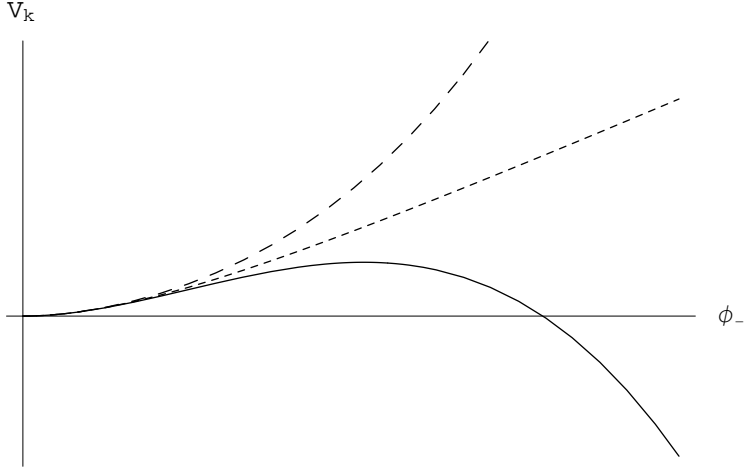}}}
\end{center}
\caption{Plot of the effective potential (\ref{exact_effective_potential_k}), with $f(\f_-)$ given by
(\ref{conformal_deformation}), on $S^3$ for $\a<\sqrt{\l}$
(long dashes), $\a=\sqrt{\l}$ (short dashes), and $\a>\sqrt{\l}$.}
\label{fig}
\end{figure}

Setting first $a=b$ in (\ref{instanton_uhp}) and letting $b\to\infty$ and $\vec{z}_o^2\to\infty$, while
keeping $b/\vec{z}_o^2=\frac{l\sqrt{\l}}{d-1}\vf_o^{\frac{2}{d-1}}$ constant, reproduces the solution
(\ref{mu=0}) with $k=0$. Moreover, setting $z^i_o=0$ and
$b=\frac{2l\sqrt{\l}}{d-1}\vf_o^{\frac{2}{d-1}}$ in (\ref{instanton_uhp}), the two choices $a=\pm\sqrt{\pm1+b^2}$
lead via the coordinate transformations given in Table \ref{slice-metrics}
to the solution (\ref{mu=0}) with $k=\pm1$ respectively. It follows that, although all three solutions
satisfy boundary conditions corresponding to the marginal deformation (\ref{conformal_deformation}),
only the solution for $k=1$ satisfies the condition $\a>\sqrt{\l}$, which is necessary in order
for the solution to be identified with an instanton. For $k=0$ instead we have $\a=\sqrt{\l}$, while
for $k=-1$, $\a<\sqrt{\l}$. This can be seen directly by looking at the extrema of the effective
potential (\ref{exact_effective_potential_k}). Namely, the equation $V'_k(\f_-)=0$ is nothing but
the gap equation $J_{f_-}=-\hp\sub{\D_+}-f'(\f_-)=0$, where now $\hp\sub{\D_+}$ is given by
(\ref{exact-mom}) and $f(\f_-)$ by (\ref{conformal_deformation}). Rearranging this equation gives
\be
k+(\l-\a^2)\left(\frac{2l}{d-1}\vf_o^\frac{2}{d-1}\right)^2=0.
\ee
Note that the VEV, $\f_o$, is undetermined for $k=0$. In the special case where this
solution is embedded into $\cn=8$ gauged supergravity in four dimensions, this is precisely the
Coulomb branch solution corresponding to $\n=-1$ in (\ref{2/3-dw}). For $k=1$ the VEV is fixed to
$\vf_o^{\frac{2}{d-1}}=1/l\tilde{l}$, which is a local {\em maximum} of the effective potential
(\ref{exact_effective_potential_k}), as can be seen from Figure \ref{fig} for the case $d=3$. This plot
also shows that the effective potential on $S^3$ is stable, marginally stable and unstable
according to whether $\a<\sqrt{\l}$, $\a=\sqrt{\l}$ and $\a>\sqrt{\l}$ respectively.

\begin{flushleft}
{\bf Vacuum decay rate}
\end{flushleft}

We have seen above that the existence of these instanton solutions for $\a>\sqrt{\l}$ coincides
with the critical point where the exact quantum effective potential (\ref{exact_effective_potential_k})
becomes unbounded from below. This suggests that the dual field theory becomes unstable and
the conformal vacuum at $\f_-=0$ decays via quantum tunneling, mediated by these instantons, to something
else \cite{dHPP}. The endpoint of this decay is unclear, however, since classical supergravity
breaks down before this endpoint is reached. In the $d=3$ case, where the system
(\ref{bulk_conformal_action}) can be embedded in $\cn=8$ gauged supergravity and the corresponding
asymptotically $AdS_4\times S^7$ solutions uplifted to 11-dimensional supergravity, this breakdown
of the supergravity description can be traced to an $S^2$ inside the $S^7$ collapsing \cite{dHPP}.
$1/N$ corrections must therefore be taken into account in order to understand the endpoint of the decay.
From the bulk point of view, the instantons signal an instability of the AdS vacuum once the modified
boundary condition (\ref{conformal_deformation}), with $\a> \sqrt{\l}$ is imposed.

The instanton solutions allow us to compute the decay probability of the conformal vacuum, which is given by \cite{Coleman:1980aw}
\be
\cp\propto\left.e^{-\G_\a[\f]}\right|_{\rm instanton},
\ee
where $\left.\G_\a[\f]\right|_{\rm instanton}$ is the effective action of the boundary theory
evaluated on the instanton VEV (\ref{VEV}). Although we do not know the general form of the effective
action, this decay rate can be computed exactly in two different ways. First, the effective
action on the VEV (\ref{VEV}) can be evaluated by computing the bulk on-shell action on the instanton
solution (\ref{instanton_uhp}), taking into account the boundary term (\ref{conformal_deformation}).
This gives
\be\label{on_shell_instanton}
\framebox[5.5in]{
\begin{minipage}{7.5in}
\begin{equation*}
\left.\G_\a\right|_{\rm instanton}=\frac{\p^{\frac{d+1}{2}}\tilde{l}^{-d}}
{\G(\frac{d+1}{2})}\frac{\a}{d}\left[1-\frac{d}{(d+1)\frac{\a}{\sqrt{\l}}
\left(\frac{\a}{\sqrt{\l}}+1\right)}F\left(\frac{d+2}{2},1;d+2;\frac{2}
{\frac{\a}{\sqrt{\l}}+1}\right)\right].
\end{equation*}
\end{minipage}
}
\ee
The second way relies on boundary quantities only. The crucial observation is that since the
value of the on-shell action does not depend on the moduli of the instanton - indeed
(\ref{on_shell_instanton}) depends only on the coupling $\l$ and the deformation parameter $\a$ -
we can go to any point in the instanton moduli space to evaluate the effective action. In particular,
we saw that there is a point in the moduli space where the instanton VEV (\ref{VEV}) is {\em constant},
namely $\f_-^{\frac{2}{d-1}}=1/l\tilde{l}$. This corresponds to the $\m=0$, $k=1$ domain wall we discussed
 above. But then the effective action reduces to the effective potential, which we have computed in (\ref{exact_effective_potential_k}).
Evaluating the effective potential on this constant VEV and multiplying by the volume of $S^d$ gives, after some manipulation using the identity
\be
F\left(-\frac d2,\frac12;1-\frac d2;-x\right)=\sqrt{1+x}-\left(\frac{d-1}{d-2}
\right)x F\left(1-\frac d2,\frac12;2-\frac d2;-x\right),
\ee
precisely (\ref{on_shell_instanton}).

\section*{Acknowledgements}

I would like to thank Jan de Boer, Sebastian de Haro, Anastasios C. Petkou, Kostas Skenderis and
J\"org Teschner for very useful comments. I also thank Ecole Polytechnique for hospitality during the completion of this work.

\appendix
\renewcommand{\thesection}{\Alph{section}}
\renewcommand{\theequation}{\Alph{section}.\arabic{equation}}

\section{The variational problem and Hamilton-Jacobi equations}
\setcounter{equation}{0}
\label{appendix}

In this appendix we consider the variational problem for the actions
(\ref{bulk_minimal_action}) and (\ref{bulk_conformal_action}) on the
regulating hypersurface $\S_r$. Specifically, we determine the
appropriate Gibbons-Hawking term and the radial canonical momenta
for each case. Moreover, we give the radial Hamiltonian densities, which, via
the Hamiltonian and momentum constraints, determine the corresponding Hamilton-Jacobi
equations.

A generic variation of the bulk action produces a bulk term proportional to
the equations of motion as well as a boundary term. Namely,
\be\label{variation}
\d S=\int_{\cm}d^{d+1}x\sqrt{g}\left(({\rm eoms})+\nabla_\m v^\m\right),
\ee
for some vector field $v^\m$. This vector field is fundamental to
the study of the variational problem and the radial Hamiltonian formalism.
For the actions (\ref{bulk_minimal_action}) and (\ref{bulk_conformal_action})
it is given respectively by
\bea\label{min-b-term}
&&v^\m_{\rm min}=-\frac{1}{\k^2}
g^{\r[\m}\nabla^{\s]}\d g_{\r\s}+\d\f\nabla^\m\f,\\
\label{conf-b-term}
&&v^\m_{\rm conf}=-\frac{1}{\k^2}\left(1-\frac{(d-1)\k^2}{4d}\f^2\right)
g^{\r[\m}\nabla^{\s]}\d g_{\r\s}-\frac{d-1}{2d}\f\d
g_{\r\s}g^{\r[\m}\nabla^{\s]}\f+\d\f\nabla^\m\f.
\eea
We can evaluate explicitly the boundary term (\ref{variation}) on
$\S_r$ using the gauge-fixed metric (\ref{gf-metric}). Generically it takes the
form
\be\label{b-term}
\cb_r\equiv \int_{\S_r}\ast v=-\d S_{GH}+\int_{\S_r}d^dx
(\d\g_{ij}\p^{ij}+\d\f \p_\f),
\ee
but the precise form of the Gibbons-Hawking term, $S_{GH}$, and of the
radial canonical momenta, $\p^{ij}$ and $\p_\f$, crucially depends
on the form of the bulk action. In Table \ref{momenta} we give
explicitly the momenta and the Gibbons-Hawking terms for minimally
and conformally coupled scalars. Note that although the Gibbons-Hawking term
for minimally coupled scalars is identical to the standard Gibbons-Hawking
term for pure gravity, this is no longer true for conformally coupled scalars.
\begin{table}
\begin{center}
\begin{tabular}{|l|l|}
\hline
\multirow{7}{*}{\rotatebox{90}{minimal}}
& \\
& $\p^{ij}=-\frac{1}{2\k^2}\sqrt{\g}(K\g^{ij}-K^{ij})$\\ & \\
& $\p_\f=\sqrt{\g}\dot{\f}$\\ & \\
& $S_{\rm GH}=-\frac{1}{\k^2}\int_{\S_r} d^dx\sqrt{\g}K$\\ & \\
\hline & \\
\multirow{6}{*}{\rotatebox{90}{conformal}}
& $\p^{ij}=-\frac{1}{2\k^2}\sqrt{\g}\left(1-\frac{(d-1)\k^2}{4d}\f^2\right)
(K\g^{ij}-K^{ij})+\frac{d-1}{4d}\sqrt{\g}\f\dot{\f}\g^{ij}$\\ & \\
& $\p_\f=\sqrt{\g}\left(\dot{\f}+\frac{d-1}{2d}K\f\right)$\\ & \\
& $S_{\rm GH}=-\frac{1}{\k^2}\int_{\S_r} d^dx\sqrt{\g}
\left(1-\frac{(d-1)\k^2}{4d}\f^2\right)K$\\ & \\
\hline
\end{tabular}
\end{center}
\caption{The canonical radial momenta and the Gibbons-Hawking terms
for the actions (\ref{bulk_minimal_action}) and (\ref{bulk_conformal_action}).
Note that $K_{ij}=\frac12\dot{\g}_{ij}$ is the extrinsic curvature of the hypersurface $\S_r$.}
\label{momenta}
\end{table}

The radial momenta given in Table \ref{momenta} are, of course, the same quantities as those one would
obtain from the functional derivatives of the off-shell bulk Lagrangian with respect to the radial
derivative of the corresponding induced field, i.e.
\be
\p^{ij}=\frac{\d L}{\d\dot{\g}_{ij}},\quad \p_\f=\frac{\d L}{\d\dot{\f}}.
\ee
However, the boundary term (\ref{b-term}) shows that they also correspond to the
functional derivatives of the the {\em regularized} on-shell action,
\be\label{reg-action}
I_r\equiv (S+S_{\rm GH})|_{\rm on-shell},
\ee
with respect to the induced fields on the hypersurface $\S_r$. Namely,
\be\label{HJ}
\p^{ij}=\frac{\d I_r}{\d\g_{ij}},\quad \p_\f=\frac{\d I_r}{\d\f}.
\ee
These relations, familiar from Hamilton-Jacobi theory, are the main reason why the radial
Hamiltonian formalism is the most direct approach for studying the supergravity limit of
the AdS/CFT correspondence. Indeed, in the simplest case of Dirichlet boundary conditions,
the AdS/CFT dictionary identifies the induced fields, e.g. $\g_{ij}$ and $\f$,
with the sources of the dual operators and the regularized on-shell action with the
generating functional of {\em regularized} connected correlation functions. It
follows that the canonical momenta given by (\ref{HJ}) correspond to the
regularized one-point functions of the dual operators with arbitrary sources.
This statement trivially carries over for {\em renormalized} correlation
functions once the covariant boundary counterterms are added to $I_r$.
Moreover, as it is extensively discussed in Section \ref{bvp}, once the covariant
boundary counterterms are added to $I_r$, one can add further appropriate
finite boundary terms in order to modify the boundary conditions.

The bulk equations of motion can be written in terms of the radial canonical momenta
using a `radial ADM formalism'. As is well known, the resulting equations are
the standard first order Hamilton equations complemented with the Hamiltonian and momentum constraints,
which reflect the diffeomorphism invariance of the theory. For the actions (\ref{bulk_minimal_action})
and (\ref{bulk_conformal_action}) the constraints take the form
\be\label{constraints}
\ch=0,\qquad 2D_i\p^i_j=\p_\f\pa_j\f,
\ee
where the Hamiltonian density, $\ch$, is given respectively by
\bea\label{Hamiltonian-min}
\ch_{\rm min}&=&\frac{2\k^2}{\sqrt{\g}}\left(\p^{ij}\p_{ij}-\frac{\p^2}{d-1}
\right)+\frac{1}{2\sqrt{\g}} \p_\f^2-\sqrt{\g}\left(-\frac{1}{2\k^2}R[\g]
+\frac12\pa^i\f\pa_i\f+V(\f)\right),\\\NO\\
\label{Hamiltonian-conf}\ch_{\rm conf}&=&\frac{2\k^2}{\sqrt{\g}}\left(1-\frac{(d-1)\k^2}{4d}\f^2
\right)^{-1}\left(\p^{ij}\p_{ij}-\frac{1}{d}\p^2\right)-\frac{2\k^2}{d(d-1)
\sqrt{\g}}\left(\p-\frac{d-1}{4}\f\p_\f\right)^2+\frac{1}{2\sqrt{\g}}\p_\f^2
\NO\\
&&-\sqrt{\g}\left(-\frac{1}{2\k^2}R[\g]+\frac{(d-1)^2}{2d(d-2)}\f^{\frac{d}
{d-1}}\D_\g\f^{\frac{d-2}{d-1}}-\frac{d(d-1)}{2\k^2l^2}+\frac{\lambda}{2}
\f^{\frac{2(d+1)}{(d-1)}}\right),
\eea
and where
\be
\D_\g\equiv -\square_\g+\frac{(d-2)}{4(d-1)}R[\g],
\ee
is the scalar conformal Laplacian in $d$ dimensions. Note that although the form of the Hamiltonian
and of the momenta is different for minimally and conformally coupled scalars, the form of
the constraints remains the same. Hamilton's equations can then be written in terms of the Hamiltonian
$H=\int d^dx\ch$ as
\bea\label{Hamilton}
&&\dot{\g}_{ij}=2K_{ij}=\frac{\d H}{\d\p^{ij}}, \quad\dot{\f}=\frac{\d H}{\d\p_\f},\\
&&\dot{\p}^{ij}=-\frac{\d H}{\d\g_{ij}}, \quad\dot{\p}_\f=-\frac{\d H}{\d\f}.
\eea
The two equations in the first line are just the inverse of the expressions in Table \ref{momenta}
for the momenta in terms of the radial derivatives of the induced fields. The two equations in the
second line give the second order equations one would obtain from the components of Einstein's
equation that are transverse to $\S_r$. However, we will not need the explicit form of these
equations since we only use the Hamilton-Jacobi formalism in this paper. This consists in
inserting the canonical momenta as derivatives of the regularized on-shell action (see (\ref{HJ}))
in the Hamilton and momentum constraints (\ref{constraints}). The resulting equations
are the Hamilton-Jacobi equations for the gravity-scalar system. Hamilton's equations
are then automatically satisfied due to the identification (\ref{HJ}).

\end{document}